\documentclass[11pt]{article}
\pdfoutput=1

\usepackage[normalem]{ulem}
\usepackage{amsmath}
\usepackage{amssymb}
\usepackage{slashed}
\usepackage{graphicx}
\usepackage{cancel}
\usepackage{float}
\usepackage{subfig}
\usepackage{graphics}
\usepackage{epsfig}
\usepackage[usenames, dvipsnames]{color}
\usepackage{xcolor}
\usepackage{soul}
\usepackage{hyperref}
\usepackage{upgreek}
\usepackage{authblk}
\usepackage{blindtext}
\usepackage{mathrsfs}
\usepackage{float}
\usepackage{boxhandler}
 \usepackage{caption}
\usepackage{morefloats}
\usepackage{mathtools} 
\usepackage{braket} 
\usepackage{pdfpages}
\usepackage{slashed}

\newcommand{\be}{\begin{equation}}
\newcommand{\ee}{\end{equation}}
\newcommand{\bea}{\begin{eqnarray}}
\newcommand{\eea}{\end{eqnarrray}}

\newcommand{\av}[1]{\langle #1 \rangle}
\newcommand{\kgnorm}[1]{( #1 )_{\mathrm{KG}}}
\newcommand{\hD}{\widehat{\Delta}}

\newcommand{\uksj}{s_{\mathbf k}}
\newcommand{\uksjn}{\tilde{s}_{\mathbf k}}
\newcommand{\ukpsj}{s_{\mathbf k'}}

\newcommand{\vkpjp}{\tilde{s}_{\mathbf k}^+}
\newcommand{\vkpjm}{\tilde{s}_{\mathbf k}^-}

\newcommand{\uqkg}{u_{\mathbf q}}
\newcommand{\uqpkg}{u_{\mathbf q'}}

\newcommand{\Aqk}{A_{\mathbf{qk}}}
\newcommand{\Bqk}{B_{\mathbf{qk}}}
\newcommand{\hP}{\widehat{\Phi}}
\newcommand{\hB}{\widehat{\Box}}
\newcommand{\wsj}{W_{\mathrm{SJ}}}
\newcommand{\kr}{\mathrm{Ker}} 
\newcommand{\im}{\mathrm{Im}} 
\newcommand{\ha}{\mathbf a } 
\newcommand{\hak}{\ha_{\mathbf {k}}}

\newcommand{\hb}{\mathbf b } 
\newcommand{\hbk}{\hb_{\mathbf {k}}}
\newcommand{\hbkd}{{\hb_{\mathbf {k}}^\dagger}}

\newcommand{\haq}{\ha_{\mathbf {q}}}
\newcommand{\haqd}{{\ha_{\mathbf {q}}^\dagger}}
\newcommand{\haqp}{\ha_{\mathbf {q'}}}
\newcommand{\haqpd}{{\ha_{\mathbf {q'}}^\dagger}}
\newcommand{\lk}{\lambda_{\mathbf{k}}}

\newcommand{\stkout}[1]{\ifmmode\text{\sout{\ensuremath{#1}}}\else\sout{#1}\fi}
\newcommand{\tre}{\text{Re}}
\newcommand{\cL}{\mathcal{L}}
\newcommand{\cM}{\mathcal{M}}
\newcommand{\Wmink}{W_{\mathrm{mink}}}
\newcommand{\phm}{m_p}
\newcommand{\tm}{m}
\usepackage{fullpage}

\begin{document}

\author[1]{Sumati Surya}

\author[1]{Nomaan X} 
\author[2]{Yasaman K. Yazdi} 
\affil[1]{\textit{Theoretical Physics Group, Raman Research Institute, C.V. Raman Avenue, Sadashivnagar, Bangalore
560 080, India}}
\affil[2]{\textit{Department of Physics, 4-181 CCIS, University of Alberta, Edmonton AB, T6G 2E1, Canada}}
\affil[ ]{\textit {ssurya@rri.res.in, nomaan@rri.res.in, kouchekz@ualberta.ca}}

\title{Studies on the SJ Vacuum in de Sitter Spacetime}

\date{}
\setcounter{Maxaffil}{0}
\renewcommand\Affilfont{\itshape\small}

\baselineskip=0.5cm

\maketitle
\begin{abstract}  
In this work we study the Sorkin-Johnston (SJ)  vacuum in de Sitter spacetime for free scalar field theory. For the massless theory we find that the  SJ vacuum can neither be obtained from the $O(4)$ Fock vacuum of Allen and Folacci nor from the non-Fock de Sitter invariant vacuum of Kirsten and Garriga. Using a causal set discretisation of a slab of 2d and 4d de Sitter spacetime,  we find the causal set SJ vacuum  for a range of masses $m \geq 0$ of the free scalar field.  While our simulations are limited to a finite volume slab of global de Sitter spacetime,  they show good convergence  as the volume is increased.  We find that the 4d causal set SJ vacuum shows a significant departure from the continuum Mottola-Allen $\alpha$-vacua.   Moreover, the causal set SJ vacuum  is well-defined for both the minimally coupled massless $m=0$ and the conformally coupled massless $m=m_c$ cases. This is at odds with earlier work on the continuum de Sitter SJ vacuum where it was argued that the continuum SJ vacuum is ill-defined for these masses. Our results hint at an important tension between the discrete and continuum behaviour of the SJ vacuum in de Sitter and suggest  that the former  cannot in general be identified with the Mottola-Allen $\alpha$-vacua even for $m>0$.   
\end{abstract} 

\section{Introduction}
As it is usually defined, the vacuum for QFT on a generic curved spacetime  relies on a choice
of observer or equivalently a choice of mode functions, and is hence non-unique. In  free scalar quantum field theory (FSQFT), the Sorkin-Johnston or SJ vacuum \cite{Johnston:2009fr, Sorkin:2011pn} is a proposal for an observer independent vacuum which {\it is} unique.  The  idea is to begin with the covariantly defined spacetime commutator or Peierls bracket 
\begin{equation}
    [\hP(x),\hP(x') ]=i\Delta(x,x'), 
\end{equation}
where the Pauli-Jordan (PJ) function $i\Delta(x,x')\equiv i \, ( \, G_R(x,x')-G_A(x,x')\,)$ and   $G_{R,A}(x,x')$  are the retarded and advanced Green functions. The  PJ function can be viewed as the integral kernel of a self-adjoint  operator $i\widehat\Delta$ on a bounded region $\mathcal{V}$ of spacetime.  Its non-zero eigenvalues thus come in positive and negative pairs, providing  a natural and covariantly defined mode decomposition into  ``SJ modes".  The positive part of the spectral  decomposition of $i\widehat\Delta$ is  then defined to be the SJ Wightman or two-point function $\wsj(x,x')$. 

It is therefore of interest to ask what new role, if any, the SJ vacuum plays in FSQFT in 
cosmologically interesting spacetimes such as  de Sitter. Using a particular limiting procedure, it was argued in \cite{Aslanbeigi:2013fga} that the SJ vacuum for global de Sitter spacetime can be  identified with  one of the known Mottola-Allen $\alpha$-vacua \cite{Mottola:1984ar,Allen:1985ux} for each value of $m^2 =\phm^2+\xi R >0$\footnote{Here $\phm$ is the physical mass. For a discussion on the meaning of mass in dS spacetime see \cite{Garidi:2003ys}.} for spacetime dimensions $d\geq 2$, {\it except} for the conformally coupled massless case $m^2=m_c^2=\frac{(d - 2)}{4(d-1)} R\equiv\xi_cR$, where the SJ vacuum was argued to be ill-defined. Since there is no known de Sitter invariant Fock vacuum for the minimally coupled massless case $m=0$ \cite{Allen:1985ux}, they also suggest that the $m=0$ SJ vacuum is ill-defined. While  general infrared considerations might be consistent with the absence of an $\tm=0$ SJ vacuum, the situation for $\tm=\tm_c$ is puzzling. 

An important subtlety in the construction of the SJ vacuum is the use of a bounded region $\mathcal{V}$ of spacetime in defining $i\widehat{\Delta}$.  This  operator is Hermitian on the space of $\mathcal{L}^2$ spacetime functions, where 
\begin{equation}
    \braket{f,g}=\int_{\mathcal{V}}dV f^*(x)g(x)
\end{equation}
defines the $\mathcal{L}^2$ inner product and $\mathcal{V}$ is a finite volume region of the full spacetime $(\mathcal{M},g)$. Thus  the SJ vacuum of $(\mathcal{M},g)$ can be obtained only in the limit $\mathcal{V}\rightarrow \mathcal{M}$.  A pertinent  question is whether the SJ  construction is sensitive to exactly {\it when} this limit is taken.  

In the literature there have been two approaches to constructing the  SJ
vacuum  arising from the choice of when to take this ``IR limit". The first and more fundamental approach is what we dub the   ``ab initio'' calculation  where the eigenfunctions and eigenvalues of $i\hD$ are obtained in the  bounded region  $\mathcal{V}$. The  SJ vacuum $\wsj(x,x')$ is obtained as the positive part of $i\hD$.   If $\wsj(x,x')$ remains well-behaved when $\mathcal{V}\rightarrow \mathcal{M}$  then this gives the SJ vacuum in $(\mathcal{M},g)$.  This is the approach followed by \cite{Afshordi:2012ez} for the massless FSQFT  in the 2d causal diamond in Minkowski spacetime. The SJ two-point function was moreover  shown to be Minkowski-like near the center of the causal diamond, with the expected 2d logarithmic behaviour. The ab initio calculation is however computationally challenging since it is non-trivial to calculate the spectral (or eigen) decomposition of $i\hD$ explicitly. Indeed, the spectral decomposition  of $i\hD$ is known in very few examples other than the 2d causal diamond \cite{Fewster:2012ew,Brum:2013bia,Buck:2016ehk,MS}. 

The second, more computationally accessible approach, which we dub the ``mode comparison" calculation,  was adopted extensively in \cite{Aslanbeigi:2013fga,Afshordi:2012jf}. The idea is to start with a  set  of  Klein Gordon (KG) modes $\{ \uqkg\}$ in  the full spacetime and  restrict them to  $\mathcal{V}$. The  SJ modes $\{ \uksj\}$ in $\mathcal{V}$ are obtained from $\{ \uqkg\}$  via a Bogoliubov transformation. 
The SJ modes are then assumed to  extend to the full spacetime only if the  coefficients of
this  transformation are well behaved in the IR limit.  
Furthermore, when the $\{ \uksj\}$ can themselves be identified with a known set of KG modes, the SJ vacuum is identified with the corresponding known KG vacuum in the full spacetime, rather than via an explicit calculation.  

In these two calculations, the IR limit is taken differently. In the former, it is taken after the finite SJ vacuum is constructed from the eigen decomposition in $\mathcal{V}$, while in the latter,  the limit is taken after the mode comparison in the full spacetime restricted to $\mathcal{V}$. In the 2d causal diamond both calculations give the same result away from the boundaries \cite{Afshordi:2012ez,Afshordi:2012jf}. However,  this is in general not guaranteed and needs to be checked case by case. The subtlety of when to take the limit was brought out in  \cite{Fewster:2012ew} for the case of ultrastatic spacetimes. There, the finite $\mathcal{V}$ SJ vacuum was shown not to be equivalent to that constructed from a Hadamard state, and in some cases, to be in an inequivalent representation altogether. However, in taking the IR limit, both  yield  the same Hadamard vacuum.   
It is the aim of this work to re-examine the de Sitter SJ vacuum from the perspective that the nature of  the SJ vacuum is sensitive to the manner in which the IR limit enters its construction. This study is significant for the definition of the SJ vacuum, since it is only if the ab initio calculation fails to survive the IR limit that we can  definitively say that there is no SJ vacuum.   

We begin with the two known $m=0$ vacua in de Sitter\footnote{There is also a de Sitter invariant and shift invariant vacuum defined in \cite{Page:2012fn}. In this paper, we do not impose shift invariance.}: the $O(4)$ invariant Fock vacuum of \cite{Allen:1987tz} and the de Sitter invariant non-Fock vacuum of \cite{Kirsten:1993ug}. In the spirit of the mode comparison calculation, we show that the SJ modes cannot be obtained via a Bogoliubov transformation from the modes that define these two vacua. The calculation is done in a symmetric $[-T,T]$ slab of global de Sitter spacetime  and the coefficients of the transformation are seen to diverge as $T\rightarrow \pi/2$ (the infinite volume limit). At present we do not have an analytic ab initio calculation of the SJ modes in de Sitter spacetime. Instead we use a causal set discretisation of a slab of de Sitter spacetime and obtain the causal set SJ vacuum via the ab initio calculation. In the massive theory in 2d, our results are in keeping with the findings of \cite{Aslanbeigi:2013fga} and agree very well with the continuum Mottola-Allen $\alpha$-vacua. On the other hand, while the $m=0$ SJ vacuum is well-defined, it appears to violate  de Sitter invariance. In the massive theory in 4d, our results show a substantial difference with the continuum expressions of \cite{Aslanbeigi:2013fga} and suggest that the causal set SJ vacuum, while de Sitter invariant, differs from the Mottola-Allen $\alpha$-vacua. For $m=0$ and $m_c$, interestingly, the SJ vacuum is well-behaved, and also does not violate de Sitter invariance. In particular, at and around $m=m_c$, the SJ vacuum behaves as a continuous function of $m$, suggesting no singular behaviour. While our numerical calculations are of course for a finite volume, by varying the IR cutoff we find  a convergence of the SJ vacuum,  which  supports our conclusions. 

In Section \ref{sjvac} we review the SJ construction, emphasising the role of the IR cutoff. In Section \ref{sec:dS} we show that the $m=0$ SJ modes in a slab of  de Sitter spacetime can neither be obtained 
from the $O(4)$-invariant Fock vacuum of \cite{Allen:1987tz} nor from the de Sitter invariant non-Fock vacuum
of \cite{Kirsten:1993ug} via a Bogoliubov transformation.  
In Section \ref{sec:cstsj} we review the causal set discretisation of de Sitter spacetime and the construction of the causal set advanced and retarded Green functions in de Sitter spacetime \cite{Nomaan:2017bpl}. 
In Section \ref{sec:numerics} we present our results from numerical simulations using a causal set discretisation of a slab of de Sitter spacetime. Our analysis begins with the massless FSQFT in 2d and 4d
causal diamonds in Minkowski spacetime. We show that the SJ vacuum looks like the Minkowski vacuum in a smaller causal diamond within the larger one, both in 2d and 4d. The former is consistent with the calculations of \cite{Afshordi:2012ez}. Next we calculate the SJ vacuum in slabs of 2d and 4d global de Sitter spacetime in the time interval $[-T,T]$ for different values of $m$. We vary $T$ as well as the density $\rho$ to look for convergence. We compare our results with the Mottola-Allen $\alpha$-vacua and show that while they agree well with the SJ vacuum (for $m>0$) in 2d, they differ significantly in 4d. We also examine the eigenvalues of the PJ operator in 2d and 4d de Sitter as a function of $m$ and find no significant changes around  $m=0$ and $m=m_c$.
In Section \ref{sec:discussion} we discuss the implications of our results. The appendices contain details of de Sitter spacetime, as well as some of the calculations required for the main text.  

In this work we have used causal sets as a covariant discretisation of the continuum. In causal set theory (CST) however, this discrete substratum is considered more fundamental than the continuum. From the CST perspective  therefore the SJ de Sitter vacuum that we have obtained {is} physically more relevant to QFT in the early universe than any  continuum vacuum. Our result that the causal set SJ vacuum differs significantly from the continuum vacua therefore suggests exciting new possibilities for CST phenomenology. An interesting future direction is to extract  observational consequences for the early universe using the causal set SJ de Sitter vacuum. 

The SJ vacuum can also be used to calculate Sorkin's spacetime entanglement 
entropy  \cite{Bombelli:1986rw,Sorkin:2012sn} both in the continuum and in a
causal set. The SJ vacuum is a pure state with zero Sorkin entanglement entropy (SEE), but its  restriction to a smaller region is not pure. In 2d Minkowski spacetime, the SEE  for a small causal diamond inside a larger one exhibits the expected logarithmic scaling behaviour with the UV cutoff \cite{Saravani:2013nwa}.  However, the calculation of the SEE for the corresponding causal set construction exhibits a spacetime
volume law scaling, unless a subtle UV double truncation is used \cite{Sorkin:2016pbz}. Since de Sitter horizons are of special interest, the causal set SJ de Sitter vacuum can be used for calculating the SEE for de Sitter horizons. In a subsequent work we will show that the double truncation procedure yields an area law for horizons in 4d de Sitter in the causal set \cite{dsxy}. 

\section{The  SJ vacuum} 
\label{sjvac}
We begin with a short introduction to the SJ vacuum construction for
FSQFT in a general globally hyperbolic, finite volume $\mathcal{V}$ region of spacetime $(\mathcal{M},g)$
\cite{Sorkin:2011pn, Aslanbeigi:2013fga, Afshordi:2012ez, Afshordi:2012jf, Sorkin:2017fcp}. 

The Klein Gordon (KG) equation in $(\mathcal{M},g)$ is  
\begin{equation} 
\biggl(\hB- m^2\biggr) \phi=0,
\end{equation} 
where $\hB \equiv g^{ab}\nabla_a \nabla_b$, and the effective mass $m^2=\phm^2 +\xi R$, where $\phm$ is the physical
mass, $R$ is the scalar curvature of $(\mathcal{M},g)$ and $\xi$ is the coupling.   
Let $\{\uqkg\}$ be a complete set of modes satisfying the KG equation in $(\mathcal{M},g)$ and orthonormal with respect to the KG symplectic form (or KG ``norm") 
\begin{equation}
\kgnorm{f,g}=\int_\Sigma (f^*\nabla_a g-g^*\nabla_a f) dS^a, 
\label{kgnorm}  
\end{equation}  
where $\Sigma$ is a Cauchy hypersurface in $(\mathcal{M},g)$. The field operator can be expressed as a mode expansion with respect to  the set $\{ \uqkg\}$    
\begin{equation}
\hP(x) \equiv \sum_{\mathbf{q}}\haq \uqkg(x) +\haqd\uqkg^*(x),
\label{field} 
\end{equation}
with $\haq,\,\haqd$
satisfying the  commutation relations
\begin{equation}
  [\haq,\haqpd]=\delta_{\mathbf{q}\mathbf{q}'},
  \quad [\haq,\haqp]=0\text{,
}\quad [\haqd,\haqpd]=0. \end{equation} 
The covariant commutation relations for the scalar field operator are
given by the Peierls bracket
\begin{equation} 
[\hP(x),\hP(x')] = i\Delta(x,x'), 
\end{equation} 
where the PJ
function is 
\begin{equation} 
i \Delta(x,x')  \equiv i( G_R(x,x')-G_A(x,x') ),
\end{equation} 
with $G_{R,A}(x,x')$ being the retarded and advanced Green
functions, respectively. In terms of the modes $\{\uqkg\}$  
\begin{equation} 
i \Delta(x,x')=\sum_{\mathbf{q}}\uqkg(x)\uqkg^*(x')-\uqkg^*(x)\uqkg(x'), 
\label{modeexp} 
\end{equation} 
and the two-point function associated with them is
\begin{equation}
    W(x,x') \equiv \sum_{\mathbf{q}}\uqkg(x)\uqkg^*(x'). 
\end{equation}
On the other hand the SJ state or equivalently the SJ two-point function $\wsj(x,x')$ for FSQFT, which as we will see below is constructed from the positive eigenspace of $i \hD$, is defined most generally by the following three conditions \cite{Sorkin:2017fcp} 
\begin{eqnarray} 
&&i\Delta(x,x') = \wsj(x,x') - \wsj(x',x),  \nonumber\\
&&\int_{\mathcal{V}} dV' \int_{\mathcal{V}} dV f^*(x') \wsj(x',x) f(x)  \geq 0,\quad(\text{Positive Semidefinite}) \nonumber \\ 
&&\int_\mathcal{V} dV' \wsj(x,x') \wsj^*(x',x'') = 0,\quad(\text{Ground state or Purity}) 
\label{eq:sjdefn}
\end{eqnarray} 
where the integrals are defined over a finite spacetime volume region $\mathcal{V}$ in the full spacetime $(\mathcal{M},g)$. 
In order to construct the SJ vacuum explicitly, the PJ
function is elevated to an integral operator in $\mathcal{V}$ 
\begin{equation}
i \hD \circ f \equiv i \int_{\mathcal{V}} \Delta(x,x')  f(x') dV_{x'}
\label{pjdef}
\end{equation} 
which acts on $\cL^2$ functions in $\mathcal{V}$ and where  
\begin{equation}
\braket{f,g}=\int_\mathcal{V} dV_x\,f^*(x)\,g(x) 
\end{equation}  
is the $\cL^2$ inner product.  
Since $\Delta(x,x')$ is  antisymmetric in its arguments, $i\hD$ is Hermitian on the space of $\cL^2$ functions in $\mathcal{V}$. Its non-zero eigenvalues, given by  
\begin{equation}
i \hD \circ \uksjn(x) =\int_{\mathcal{V}} dV_{x'}\,i\Delta(x,x')\uksjn(x')=\lk \uksjn(x)
\label{eq:eigen}
\end{equation} 
therefore come in pairs $(\lk, -\lk)$, corresponding to the
eigenfunctions $(\vkpjp, \vkpjm)$ where $\vkpjm=(\vkpjp)^*$.\footnote{We adopt the notation that the $\tilde{s}_k$ are the un-normalised (with respect to the $\mathcal{L}^2$ norm) SJ eigenfunctions,  whereas the $s_k$ without the tilde are the normalised SJ eigenfunctions.} This is
the central eigenvalue problem in the ab initio calculation of the SJ vacuum.  

It was shown in \cite{Sorkin:2017fcp} that 
\begin{equation} 
\kr (\hB -\phm^2) = \overline{\im (\hD)}, 
\label{eq:kerim} 
\end{equation} 
where the operators are defined in $\mathcal{V}$\footnote{In a spacetime of constant scalar curvature, $m$ defined above is constant, and hence this result continues to hold when $\phm$ is replaced by $m$.}. This means that the 
eigenvectors in the image of $i \hD$ (i.e., excluding those in $\kr (i\hD)$) span the full solution space of the
KG operator. One therefore has an {\sl intrinsic} and
coordinate independent separation of the space of solutions into the 
positive and negative eigenmodes of $i\hD$.\footnote{This is not unlike the polarisation in geometric quantisation.} 
The field operator thus has a 
{\it coordinate invariant} or observer independent decomposition 
\begin{equation}
\hP(x)=\sum_{\mathbf{k}}\hbk
\uksj(x)+\hbkd \uksj^*(x)
\label{field} ,
\end{equation}
where the SJ vacuum state is defined as 
\begin{equation} 
\hbk\ket{0_{SJ}}=0 \quad \forall\,\mathbf{k},
\end{equation} and  
\begin{equation} 
\uksj = \sqrt{\lk} \vkpjp   
\end{equation}  
are the normalised {\sl SJ modes}\footnote{For dimensional considerations, see Appendix D.}  which form an orthonormal set in $\overline{\im(i\hD)}$ with respect to the
$\cL^2$ norm
\begin{eqnarray}
\braket{\uksj,\ukpsj}&=&{\lk}\delta_{\mathbf{k}\mathbf{k}'} \nonumber \\ 
\braket{\uksj^*,\ukpsj}&=&0.
\end{eqnarray}
Using the spectral decomposition 
\begin{equation} 
i \Delta(x,x') = \sum_{\mathbf k}  \uksj(x) \uksj^*(x') -
\uksj^*(x) \uksj(x') , 
\label{sjmodeexp}
\end{equation}
the SJ two-point function in $\mathcal{V}$ is the positive part of $i\hD$ 
\begin{equation} 
\wsj(x,x') \equiv  \sum_{\mathbf k} \uksj(x) \uksj^*(x'). 
\label{sjvacuum} 
\end{equation} 
If $\wsj(x,x')$ remains well-defined as the IR cutoff is taken to infinity, this defines the SJ vacuum in the full spacetime $(\cM,g)$. The SJ construction from the eigenvalue problem \eqref{eq:eigen} through to \eqref{sjvacuum} is the ab intio calculation referred to in the introduction. 

Alternatively, one can also obtain the SJ modes via a mode comparison calculation. Given the equality in \eqref{eq:kerim} between $\overline{\im(\hD)}$ and the
KG solution space, there must  exist a transformation between the KG modes $\{ \uqkg\}$ in $\mathcal{V}$ and
the SJ modes $\{\uksj\}$, even though the former need not be orthonormal with respect to the $\cL^2$ inner product.  Let 
\begin{equation}
\uksj(x)=\sum_{\mathbf{q}}\uqkg(x)\Aqk+\uqkg^*(x)\Bqk,
\label{uksjuqkg}
\end{equation}
where $\Aqk=\kgnorm{\uqkg,\uksj}$ and $\Bqk =\kgnorm{\uqkg^*,\uksj}$. Further, if we act with $i\Delta$ on \eqref{uksjuqkg} and use \eqref{modeexp}, we can also write $\Aqk= \dfrac{1}{\lambda_{\mathbf{k}}}\av{\uqkg,\uksj}$ and $\Bqk =-\dfrac{1}{\lambda_{\mathbf{k}}}\av{\uqkg^*,\uksj}$.
Using the fact that \eqref{sjmodeexp} and \eqref{modeexp} must be equal, we get the algebraic relations 
\begin{eqnarray}
\sum_{\mathbf{q}}\mathbf{A_{qk'}}\Aqk^*-\mathbf{B_{qk'}}\Bqk^*&=&\delta_{\mathbf{k}\mathbf{k}'}\nonumber\\
\sum_{\mathbf{q}}\mathbf{B_{qk'}} \Aqk -\mathbf{A_{qk'}} \Bqk &=&0.
\label{ABrelations} 
\end{eqnarray}
Additionally, if the KG modes themselves satisfy the $\cL^2$ orthonormality condition 
\begin{equation}
    \langle \uqkg,\uqpkg\rangle=\delta_{qq'},\indent \langle \uqkg^*,\uqpkg \rangle=0,  
\end{equation}
then the above equations simplify considerably as shown in \cite{Afshordi:2012jf}.\footnote{In assuming a  discrete index $\mathbf q$ we are already working in a bounded region of spacetime.} 
It is important to note that since the $\cL^2$ norm is defined for finite $\mathcal{V}$, the above calculations are limited to finite $\mathcal{V}$. Moreover, there are potential subtleties in identifying $\kr (\hB -m^2)$ in $\mathcal{V}$, starting from the solutions in the full spacetime.  

The question of course is whether the limits involved in the first and second approaches (that is, whether finding the SJ modes before or after taking the infrared limit) commute. 
A case in point is the 2d causal diamond in Minkowski spacetime where the SJ modes for the
massless scalar field are not simply linear combinations of plane
waves, but also include an important $\mathbf k$ dependent constant
\cite{Afshordi:2012ez, Johnston:2010su}, which {\it is} a solution for finite
$\mathcal{V}$. The two sets of eigenfunctions of $i\Delta$ are  
\begin{eqnarray} 
f_k(u, v) &= & e^{iku} - e^{ikv} \label{fmode}\\ 
g_k(u, v) &=& e^{iku} + e^{ikv} - 2\cos kL\label{gmode}, 
\end{eqnarray} 
where $u$ and $v$ are lightcone coordinates, and $2 L$ is the side length of the diamond. The eigenvalues are $\lk=L/k$ for both sets. For the $f$-modes,
$k$ is $k=n\pi/L$ with $n=\pm 1, \pm 2,...$ while for the $g$-modes $k$ satisfies the condition $\tan(kL)=2 kL$.       
In order to make contact with the IR limit, $W(x,x')$ was studied in a small region in the interior of the larger diamond, which to leading order was found to have the form of the (IR-regulated) 2d  Minkowski vacuum \cite{Afshordi:2012ez}. A similar conclusion was reached in \cite{Afshordi:2012jf} using the Bogoliubov prescription, and hence in this simple example, the results seem to be independent of the limiting procedure.  

\section{The massless de Sitter SJ vacuum } 
\label{sec:dS} 
In \cite{Aslanbeigi:2013fga} the mode comparison calculation was used to find the SJ modes in de Sitter spacetime. A  
restriction of  the Euclidean modes \cite{Chernikov} (which themselves are one of the $\alpha$-modes) in global de Sitter to a finite slab $\mathcal{V}$ was used as the starting point. Assuming that these modes are complete in  $\kr (\hB -m^2)$ when restricted to $\mathcal{V}$, they  solve \eqref{ABrelations} to get the SJ modes $\{ \uksj\}$, \eqref{uksjuqkg}. These can in turn be  identified with one of the
other (restricted to $\mathcal{V}$) $\alpha$-modes depending on the value of  $m$,  and thence the SJ vacuum is identified with the corresponding $ \alpha$-vacuum in the IR
limit for each $m$. Surprisingly, however, this identification fails in the conformally coupled massless case, $m_c=\frac{(d-2)}{4(d-1)} R$, since the Bogoliubov transformation breaks down. For this
and the minimally coupled massless case, $m=0$ (for which there is no $ \alpha$-vacuum), it is suggested that the SJ prescription itself
breaks down and that there is no de Sitter SJ vacuum.  
In both these cases however,  the
SJ modes must be well-defined when there is a finite $T$ IR cutoff. Strictly, it is only if an ab initio calculation
 of the SJ two-point functions fails to survive the IR limit that we can state that there is no SJ vacuum. 
 
The KG modes for the massive scalar field in global de Sitter are the Mottola-Allen $\alpha$-modes
which include the Euclidean modes as a special case. The mimimally coupled massless scalar field is
known not to admit a de Sitter invariant Fock vacuum (Allen's theorem)
\cite{Allen:1985ux}. We note here that the proof of this theorem relies heavily on the use of the KG inner product.  

A question that poses itself then is: if an SJ vacuum for $m=0$ did exist, would it violate de Sitter invariance
or the Fock condition? This question cannot be answered using Allen's theorem, because it does  not apply to the SJ construction due to its use of the $\cL^2$ inner product.
Starting with a Fock vacuum defined with respect to an orthonormal basis $\{\phi_n(x)\}$ of the solution space of the KG equation, Allen shows that for the $m=0$ case the symmetric two-point function defined by
\begin{equation}
    G^{(1)}_\lambda(x,x')=\langle\lambda|\Phi(x)\Phi(x')|\lambda\rangle=\sum_n\,\phi_n(x)\phi^*_n(x')+\phi^*_n(x)\phi_n(x')
    \label{g1sum}
\end{equation}
    must satisfy 
    \begin{equation}
        G^{(1)}(x,x')+G^{(1)}(x,\Bar{x}')\neq C \quad\text{everywhere}
        \label{allencondition}
    \end{equation}
     for some $C\in \mathbb{R}$, where $\Bar{x}'$ represents the antipodal point of $x'$. In \cite{Allen:1985ux} the de Sitter invariant $G^{(1)}(x,x')$ fails to satisfy the required condition \eqref{allencondition}, leading to the conclusion that 
     the assumption that it is a Fock vacuum is false. Importantly the proof of condition \eqref{allencondition} relies on the use of the KG inner product and it no longer holds when we use the $\cL^2$ inner product for the vacuum state construction.\footnote{The use of the $\cL^2$ inner product  for the SJ modes suggests the possibility  that the SJ vacuum exists in a different sector of the theory.}
     
     It is also worth mentioning at this point that because the $\cL^2$ inner product is only defined in a finite region of spacetime\footnote{Allen's theorem continues to hold in a finite region of spacetime as long as we choose this region to be symmetric about $\tau=0$, where $\tau$ is the time in hyperbolic coordinates \eqref{globmetric}.}, the entire prescription inherently breaks de Sitter invariance. In the case of global de Sitter with an IR cutoff at $[-T, T]$, this is certainly the case. Since the spatial part is compact we manage to preserve $O(4)$ invariance. However, the idea is, as in \cite{Aslanbeigi:2013fga}, to take the temporal cutoffs to infinity\footnote{In the causal set case we cannot take these temporal cutoffs to infinity, but we try to reach an asymptotic regime.} and make statements that have full de Sitter invariance. 
     
     On the other hand, as in the 2d diamond,
one might imagine that away from the boundaries, there is an
approximate isometry that is retained. However, 
even if the PJ operator is itself approximately invariant,  this does
not imply that the two-point
function is, since the latter is simply the positive part of the PJ operator. It is only if the isometries preserve the positive and
negative eigenspaces separately that this can be the
case. 

Let us address this question by asking if the known de Sitter violating
vacuum, the so-called $O(4)$ vacuum \cite{Allen:1987tz} is related to the SJ vacuum via a Bogoliubov transformation as in \cite{Aslanbeigi:2013fga}.  We work in the conformal coordinates \eqref{confmetric2}
\begin{equation}
ds^2=\frac{1}{H^2\sin^2\eta}[-d\eta^2+d\Omega^2(\chi,\theta,\phi)\,]\label{confmetric1},
\end{equation}
where we have shifted $\tilde{T}\rightarrow\eta=\tilde{T}+\pi/2$ so that $\eta\in[0,\pi]$ and $(\chi,\theta,\phi)$ are coordinates on $S^3$.
The  $O(4)$ modes are 
\begin{equation} 
u_{klm}(x)=HX_k(\eta)Y_{klm}(\chi,\theta,\phi) ,
\end{equation} 
where $k=0,1,...;\,l=0,1...k;\,m=-l,-l+1,...l-1,l$. For $k=0$,  
\begin{equation} 
X_0(\eta)=A_0\bigg(\eta-\frac{1}{2}\sin2\eta-\frac{\pi}{2}\bigg)+B_0, 
\end{equation} 
and for $k\neq 0$ 
\begin{equation} 
X_k(\eta)=\sin^{3/2}(\eta)(A_kP^{3/2}_{k+1/2}(-\cos\eta)+B_kQ^{3/2}_{k+1/2}(-\cos\eta)),   
\label{xk}
\end{equation} 
where $P^\mu_\nu(x),\,Q^\mu_\nu(x)$ are independent, associated Legendre
functions defined for real $x\in[-1,1]$ as in \cite{Gradshteyn}:  
\begin{eqnarray}
P^\mu_\nu(x)=\bigg(\frac{1+x}{1-x}\bigg)^{\mu/2}\frac{_2F_1(-\nu,\nu+1,1-\mu;(1-x)/2)}{\Gamma(1-\mu)},\\
Q^\mu_\nu(x)=\frac{\pi}{2\sin\mu\pi}\bigg(P^\mu_\nu(x)\cos\mu\pi-\frac{\Gamma(\nu+\mu+1)}{\Gamma(\nu-\mu+1)}P^{-\mu}_\nu(x)\bigg).
\end{eqnarray}    
Note that the $k\neq0$ modes are the same as the  Euclidean modes. The $Y_{klm}$ are spherical harmonics that satisfy 
\begin{equation}
\int d\Omega(\chi,\theta,\phi)Y_{klm}Y^*_{k'l'm'}=\delta_{kk'}\delta_{ll'}\delta_{mm'}.
\end{equation}
The coefficients for $k=0$ are $A_0=-i\alpha,\,B_0=(1/4+i\beta)/\alpha$, where
$\alpha,\beta\in\mathbb{R}$. The coefficients for $ k\neq 0$ are
\begin{equation} 
 A_k=\bigg(\dfrac{-1+i}{\sqrt{2}}\bigg)\sqrt{\dfrac{\pi}{4k(k+1)(k+2)}},
 \quad B_k=\dfrac{-2i}{\pi}A_k.
 \end{equation}

These $O(4)$ modes are orthonormal with respect to the KG inner product but as mentioned in the last section, the Bogoliubov coefficients are defined by their $\cL^2$ inner products so we must evaluate these. We also need a choice of the finite spacetime region $\mathcal{V}$ for the $\cL^2$ inner product, we consider a slab of dS spacetime such that $\eta\in(a,b)$, the infinite volume limit corresponds to $a\rightarrow 0,\,b\rightarrow\pi$. We have  
\begin{eqnarray}
\braket{u_{klm},u_{k'l'm'}}&=&H^2\int dV_x\,X^*_k(\eta)X_{k'}(\eta)Y^*_{klm}Y_{k'l'm'}\nonumber\\
&=&\frac{1}{H^2}\delta_{kk'}\delta_{ll'}\delta_{mm'}\int_a^b\frac{d\eta}{\sin^4\eta}X^*_k(\eta)X_{k}(\eta)\nonumber\\
&=&\delta_{kk'}\delta_{ll'}\delta_{mm'}T_k,
\label{tko4}
\end{eqnarray}
\begin{eqnarray}
\braket{u^*_{klm},u_{k'l'm'}}&=&\frac{(-1)^k}{H^2}\delta_{kk'}\delta_{ll'}\delta_{mm'}\int_a^b\frac{d\eta}{\sin^4\eta}(X_k(\eta))^2\nonumber\\
&=&\delta_{kk'}\delta_{ll'}\delta_{mm'}D_k.
\label{dko4}
\end{eqnarray}
The factor $(-1)^k$ in the second expression  is due to the choice of
spherical harmonics with the special property
$Y^*_{klm}=(-1)^kY_{klm}$ \cite{Aslanbeigi:2013fga}. These equations define $T_k$ and $D_k$ ($T_k$ is real by definition). Also note that $T_k$ and $D_k$ will necessarily blow up in the infinite volume limit.   

The Bogoliubov coefficients to obtain the SJ modes \eqref{uksjuqkg} from these $O(4)$ modes simplify to 
\begin{eqnarray}
A_{qk}&=&\dfrac{1}{\lambda_{k}}\sum_{n}\left(\delta_{qn}T_qA_{nk}+\delta_{qn}D_q^*B_{nk}\right)
          = \dfrac{1}{\lambda_{k}}(T_qA_{qk}+D_q^*B_{qk})\nonumber \\
B_{qk}&=&-\dfrac{1}{\lambda_{k}}\sum_{n}\left(\delta_{qn}D_qA_{nk}+\delta_{qn}T_qB_{nk}\right)=-\dfrac{1}{\lambda_{k}}(D_qA_{qk}+T_qB_{qk}),
\label{redAB}
\end{eqnarray}  
where the index $q$  implicitly contains the $l$ and $m$ indices and $\delta_{ll'}, \delta_{mm'}$ are omitted from the expressions. Inserting these expressions into \eqref{ABrelations} we find that 
\begin{eqnarray}
\sum_q\{(T_q^2-|D_q|^2)(A_{qk'}A_{qk}^*-B_{qk'}B_{qk}^*)\}&=&\lambda_k^2\delta_{kk'}
\nonumber  \\ 
\sum_q\{(T_q^2-|D_q|^2)(A_{qk}B_{qk'}-A_{qk'}B_{qk})\}&=&0. 
\label{simpleAB}
\end{eqnarray}
A convenient parameterisation is 
\begin{equation}
A_{qk}=\delta_{qk}\cosh\alpha_k,\quad\quad
B_{qk}=\delta_{qk}\sinh\alpha_k\,e^{i\beta_k}. 
\end{equation} 
From \eqref{simpleAB} this gives 
\begin{equation}
\lambda_k=\sqrt{T_k^2-|D_k|^2},
\label{lamds}
\end{equation}
which along with  \eqref{redAB} implies that 
\begin{eqnarray}
\lambda_k\cosh\alpha_k&=&T_k\cosh\alpha_k+D_k^*\sinh\alpha_k\,e^{i\beta_k}\nonumber\\
\text{or}\quad\tanh\alpha_k\,e^{i\beta_k}&=&\frac{\lambda_k-T_k}{D^*_k}=\frac{T_k-\lambda_k}{|D_k|}\,e^{i(\arg\,D_k+\pi)}. 
\end{eqnarray}
Defining $r_k\equiv \dfrac{D_k}{T_k}$, we see after some algebra and use of the double angle formula for $\tanh$ that  $\beta_k=\arg
r_k+\pi$ and  $\alpha_k=\frac{1}{2}\tanh^{-1}|r_k|$. Thus the Bogoliubov coefficients depend (via $\alpha_k$ and $\beta_k$) only on $r_k$, which can be finite in the infinite volume limit \textit{even if} $T_k$ and $D_k$ diverge. Note that if $|r_k|=1$, $\alpha_k$ and therefore the Bogoliubov coefficients 
diverge. When this happens the SJ vacuum cannot be obtained through a Bogoliubov transformation.

 From \eqref{tko4} and \eqref{dko4} one can see that the Bogoliubov transformation does not mix different $k$'s. In particular, it does not mix $k\neq0$ modes with the $k=0$ mode. We already know from \cite{Aslanbeigi:2013fga} that the Euclidean modes (which are the same as the $O(4)$ modes for $k\neq0$) do not admit a well-defined Bogoliubov transformation to the SJ modes ($|r_k|=1$ for these modes) in the infinite volume limit. It immediately follows that the transformation from the $O(4)$ modes to the corresponding SJ state is ill-defined, and an SJ state with $O(4)$ symmetry cannot be derived in this way. In Appendix B we calculate these transformations explicitly. We also present the $k=0$ transformation which turns out to be the only well-defined one.

In a similar manner, we also find that the modes that define the non-Fock but de Sitter invariant vacuum of Kirsten and Garriga \cite{Kirsten:1993ug} 
are unable to produce an SJ vacuum via the mode comparison method. The Kirsten and Garriga modes are closely related to the $O(4)$ modes, and in fact are identical to them for $k\neq 0$. For $k=0$, we have
\begin{equation}
    X_0=\frac{H}{\sqrt{2}}\left[Q+\left(\eta-\frac{1}{2}\sin 2\eta-\frac{\pi}{2}\right)P\right].
\end{equation}
We use the same notation as in \cite{Kirsten:1993ug}. The coefficients of $Q$ and $P$ are solutions to the field equation that satisfy the following commutation relations
\begin{equation}
\left[Q,P\right]=i,\indent \left[\hak,Q\right]=\left[\hak,P\right]=0,
\end{equation}
where $\hak$ are the annihilation operators associated to the $k\neq 0$ modes. The details of the transformation between the Kirsten and Garriga modes and the SJ modes are presented in Appendix C. Again, we find that the $k=0$ transformation is the only well-defined one.

\section{The SJ vacuum on the Causal Set} 
\label{sec:cstsj} 

While there is  progress on finding the SJ modes via an ab initio 
calculation in some 2d as well as higher dimensional examples
 \cite{Buck:2016ehk,MS}, the calculation in 
global de Sitter is considerably more difficult. In the absence of this, we
can still carry out numerical calculations\footnote{The bulk of the simulations for this work were done using \texttt{Mathematica} \cite{Mathematica}.} using causal sets to study the two-point
function. Causal sets are not only a natural covariant discretisation
of the continuum, but also may contain important signatures of quantum
spacetime. This makes the ab initio results in the causal set even more interesting than the ab initio results in the continuum.

We begin this section with laying out some basic properties of CST. 
\subsection{Causal Sets and Sprinkling}
A \textit{causal set} $\mathcal{C}$ is a set together with an order-relation $\preceq$ that  $\forall\,x,y,z\in\mathcal{C}$ satisfies the following conditions:
\begin{enumerate}
    \item \textit{Reflexivity:} $x\preceq x$
    \item \textit{Antisymmetry:} $x\preceq y\preceq x\Rightarrow x=y$
    \item \textit{Transitivity:} $x\preceq y\preceq z\Rightarrow x\preceq z$
    \item \textit{Local finiteness:} $|\{z\in\mathcal{C}|x\preceq z\preceq y\}|<\infty$
\end{enumerate}
Here $|\cdot|$ denotes the cardinality of a set. The elements of $\mathcal{C}$ are spacetime events and the order-relation $\preceq$ denotes the causal order between the events. If $x\preceq y$ we say ``$x$ causally precedes $y$", and we write $x\prec y$ if $x\preceq y$ and $x\neq y$. Causal relations on a Lorentzian manifold (without closed timelike curves) obey conditions 1-3. Condition 4 ensures that there are a finite number of events in any causal interval; this brings in discreteness. 

Two useful ways of characterizing a causal set are the \textit{causal matrix} $C$ and the \textit{link matrix} $L$ defined as 

  \[  C_{xy}:=                                        
\left\{                                                          
	\begin{array}{ll}
		1  & \mbox{if } y \prec x \\
		0 & \mbox{} \text{otherwise}
	\end{array},
\right.
\quad\quad
L_{xy}:=                                        
\left\{                                                          
	\begin{array}{ll}
		1  & \mbox{if } y \prec x \,\text{and}\, |(x,y)|=0\\
		0 & \mbox{} \text{otherwise}
	\end{array}
\label{cl},
\right.
\]

where $(x,y)$ is the set of points that lie in the causal interval between $x$ and $y$, and the subscript $xy$ refers to indices corresponding to elements $x$ and $y$. We refer the reader to \cite{Bombelli:1987aa,Surya:2011yh,Dowker:2005tz} for more details on CST.

\textit{Sprinkling} is the process of picking points randomly from a region of spacetime $(\mathcal{M},g)$ with a given constant density $\rho$. This generates a causal set corresponding to $(\mathcal{M},g)$. The number of points picked in each realisation follows a Poisson process whose mean depends on the spacetime volume of the region. The causal ordering is inherited from the region's causal ordering restricted to the sprinkled points. The causal sets so obtained are said to  approximate  $(\mathcal{M},g)$. 

Sprinkling into regions of Minkowski spacetime has been discussed elsewhere (see e.g. \cite{Johnston:2010su}). Here we briefly describe the process for de Sitter spacetime. 

A convenient coordinate system in which to do the sprinkling for de Sitter is  the conformal coordinate system of \eqref{confmetric2}. This allows us to work with the simpler conformally related  metric in analyzing the causal structure of de Sitter spacetime.
The sprinkling can be done in two steps. In the first step we pick points randomly on the spatial part, i.e., the sphere $S^{d-1}$. One simple way (by no means unique) to do this is to generate normalised $d$-dimensional vectors. These will automatically lie on the surface of $S^{d-1}$. The corresponding spherical coordinates can be obtained by using the standard Cartesian to spherical coordinate transformation.

\begin{figure}[H]
    \centering
    \subfloat[Conformal coordinates]{\includegraphics[width=.38\linewidth]{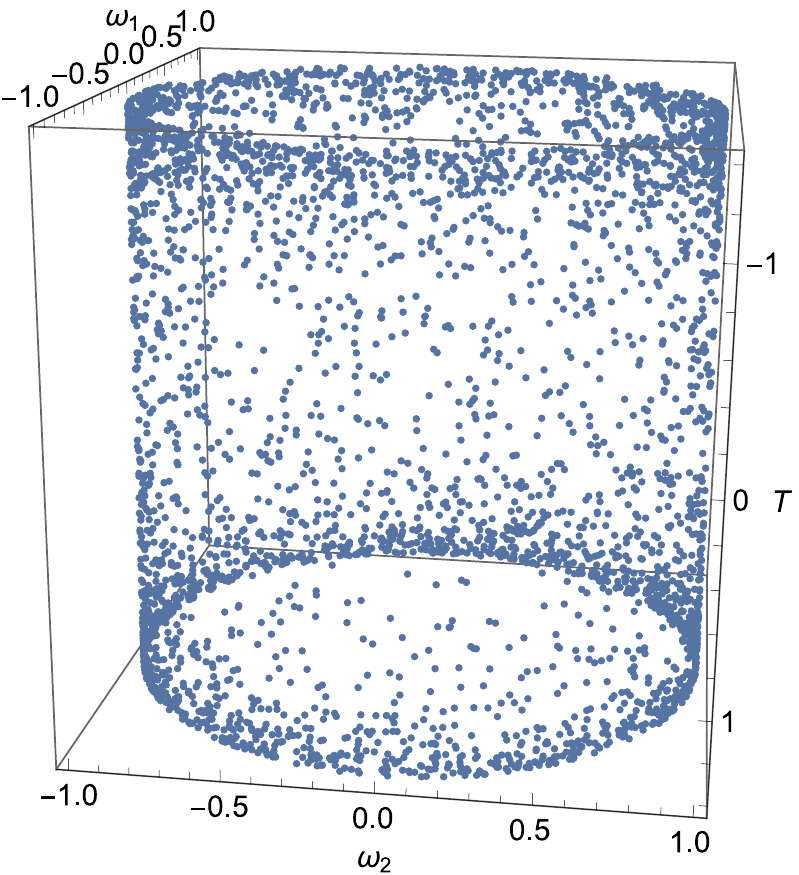}}
    \subfloat[Global coordinates]{\includegraphics[width=.40\linewidth]{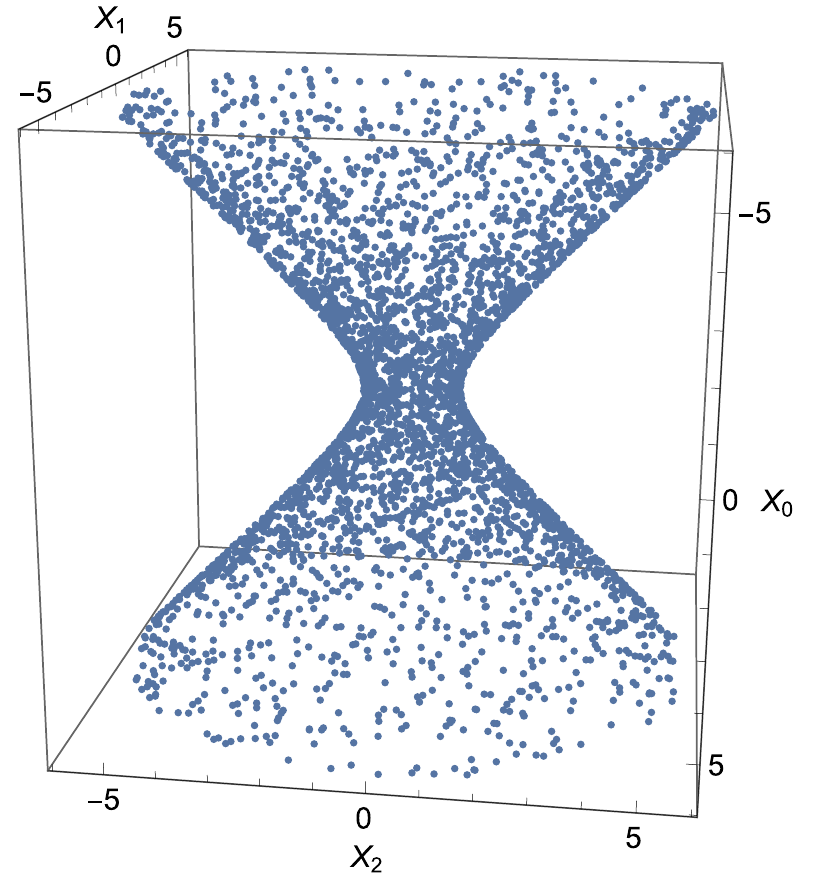}}
    \caption{A sprinkling of $N=4000$ elements for the time interval $-1.4<\tilde{T}<1.4$, $\ell=1$.}
    \label{dssprinkling}
\end{figure}

In the second step we need to obtain the temporal part of the coordinates. As is evident from the  metric, this isn't uniformly distributed but depends on the conformal factor. The effect of the conformal factor can be incorporated by defining a normalised probability distribution with a probability density function equal to $(H\cos T)^{-d}$ in the region of interest. Picking points from this distribution will give us the temporal part of the coordinates. Combining the coordinates from the two steps, we have the required sprinkling. A typical sprinkling is shown in figure \ref{dssprinkling}.  

\subsection{Green Functions}
\label{subsec:green}
The SJ vacuum is constructed from the advanced and retarded
Green functions. In \cite{Johnston:2008za} these were constructed for causal sets that approximate causal intervals\footnote{These are also known as \textit{causal diamonds} or \textit{Alexandrov intervals}.} in $2$d
and $4$d Minkowski spacetime. In \cite{Nomaan:2017bpl} it was shown that the same
construction can be extended to a larger class of spacetimes,
including de Sitter. These results are briefly summarised here.    

The massive retarded Green function in a
globally hyperbolic  $d$-dimensional spacetime $(\mathcal{M}, g)$,
satisfies
\begin{equation}
(\Box_x - m^2) G_m(x,x')= - \frac{1}{\sqrt{-g(x)} }\delta(x-x')\,.
\label{kgeq}
\end{equation}
It can also be written as
\begin{eqnarray}
\label{conv}
G_m&=&G_0 - m^2\,G_0*G_0 + m^4 \,G_0*G_0*G_0+ \ldots = \sum_{k=0}^\infty(-m^2)^k \underbrace{G_0 * G_0* \ldots G_0}_{k+1}\nonumber\\
&=&G_0 - m^2 G_0*G_m\,,
\end{eqnarray}
where $G_0$ is the massless retarded Green function satisfying \eqref{kgeq} with $m=0$. The convolution $A*B$ is defined as
\begin{equation}
(A\ast B)(x,x')\equiv \int d^dx_1 \sqrt{-g(x_1)} A(x,x_1) B(x_1,x')\,.
\end{equation}
Once we have $G_0$, then, we can write down a formal series
for $G_m$.

On a causal set of size $N$ and density $\rho$, if we have an analog of the massless retarded Green function, $K_0(x,x')$, we can propose a massive retarded Green function
$K_m(x,x')$ via the replacement
\begin{equation}
\int d^d x \rightarrow  \rho^{-1} \sum_{\textrm{causal set elements} }\,,
\end{equation}
leading to
\begin{equation}
\label{convK}
K_m =  K_0- \frac{m^2}{\rho} K_0\boldsymbol{\cdot} K_0+\frac{m^4}{\rho^2} K_0\boldsymbol{\cdot}K_0\boldsymbol{\cdot}K_0 + \ldots = K_0 - \frac{m^2}{\rho} K_0\boldsymbol{\cdot}K_m,
\end{equation}
where the convolutions have become dot products of $N\times N$ matrices. The series terminates and is well-defined for each pair of elements. We can rewrite the above equation in a compact form as
\begin{equation}
    K_m=K_0 \bigg(\mathbb{I}+\frac{m^2}{\rho} K_0\bigg)^{-1}\label{retgreen},
\end{equation}
where $\mathbb{I}$ is the $N\times N$ identity matrix. To establish a correspondence with the retarded Green function in the continuum, we need to average over multiple sprinklings (with the same density) of the causal set and then take the limit $\rho\rightarrow\infty$ i.e.
\be
G_m(x,x')=\lim_{\rho\rightarrow\infty}\av{K_m(x,x')}.
\ee
In our analysis, due to computational limitations, we use single realisations of the causal set and hence we also use the standard error of the mean (SEM) instead of the standard deviation as an estimate of error. 

In \cite{Johnston:2008za} it was shown that for 2d and 4d Minkowski spacetime, 
\begin{equation}
     K_0(x,x'):=                                        
\left\{                                                          
	\begin{array}{ll}
		\dfrac{1}{2}\,C(x,x')  &  \mbox{} d=2 \\
		\dfrac{\sqrt{\rho}}{2\pi\sqrt{6}}L(x,x') & \mbox{} d=4
	\end{array}
\right.
\label{masslessgreen}
\end{equation}
and 
\begin{equation}
     K_m(x,x'):=                                        
\left\{                                                          
	\begin{array}{ll}
		\dfrac{1}{2}\,C(x,x') \bigg(\mathbb{I}+\dfrac{m^2}{\rho} C(x,x')\bigg)^{-1} &  \mbox{} d=2 \\
		\dfrac{\sqrt{\rho}}{2\pi\sqrt{6}}L(x,x')\bigg(\mathbb{I}+	\dfrac{m^2}{2\pi\sqrt{6\rho}}L(x,x')\bigg)^{-1} & \mbox{} d=4
	\end{array}
\right.
\label{massivegreen}
\end{equation}
are good causal set analogs of the corresponding massless and massive retarded Green functions in the continuum. For comparison, the corresponding continuum retarded Green functions are
\begin{equation}
     G_0(x,x'):=                                        
\left\{                                                          
	\begin{array}{ll}
		\frac{1}{2}\,\theta(t-t')\theta(\tau^2(x,x'))  &  \mbox{} d=2 \\
		\frac{1}{2}\,\theta(t-t')\theta(\tau^2(x,x'))\dfrac{1}{2\pi}\delta(\tau^2(x,x'))& \mbox{} d=4
	\end{array}
\right.
\label{masslessgreencont}
\end{equation}
and 
\begin{equation}
     G_m(x,x'):=                                        
\left\{                                                          
	\begin{array}{ll}			\dfrac{1}{2}\,\theta(t-t')\theta(\tau^2(x,x'))J_0(m\tau(x,x'))  &  \mbox{} d=2 \\
		\dfrac{1}{2}\,\theta(t-t')\theta(\tau^2(x,x'))\bigg(\dfrac{1}{2\pi}\delta(\tau^2(x,x'))-\dfrac{m}{4\pi}\dfrac{J_1(m\tau(x,x'))}{\tau(x,x')} \bigg) & \mbox{} d=4
	\end{array}
\right.,
\label{massivegreencont}
\end{equation}
where $\tau(x,x')$ is the proper time between $x$ and $x'$, and $J_\alpha$ is a Bessel function of the first kind of order $\alpha$.

The expectation values of the causal set expressions \eqref{masslessgreen}  are
\begin{equation}
     \av{K_0(x,x')}:=                 
\left\{                               
	\begin{array}{ll}
		\dfrac{1}{2}\theta(t-t')\theta(\tau^2(x,x'))  &  \mbox{} d=2 \\
		\dfrac{\sqrt{\rho}}{2\pi\sqrt{6}}\exp(-\rho\,V(x,x')) & \mbox{} d=4
	\end{array}
\right.
\label{masslessgreenexp},
\end{equation}
where $V(x,x')$ is the spacetime volume of the causal interval between $x$ and $x'$. We can see by comparing the expressions above that in $2d$, $\av{K_0(x,x')}$ gives the continuum retarded Green function even without taking the limit $\rho\rightarrow\infty$.  This is not the case for $m\neq0$ in $2d$ or for any mass in $4d$.

In the case of de Sitter spacetime, it was shown in \cite{Nomaan:2017bpl} that the argument leading to \eqref{retgreen} can be used with a modified mass term $m'^2=\phm^2+(\xi-\xi_c)R=m^2-\xi_cR=m^2-m_c^2$. This is possible because the scalar curvature $R$ is a constant in de Sitter spacetime. The causal set massless retarded Green functions  given in \eqref{masslessgreen} also carry over to de Sitter spacetime, where they correspond to the $m_c$ case. Therefore starting from these, we can obtain the retarded Green functions for other masses and arbitrary couplings  using 
\be
  K_{m}=K_{m_c} \bigg(\mathbb{I}+\frac{1}{\rho}(m^2-m_c^2) K_{m_c}\bigg)^{-1}\label{retgreendS}.
\ee

In our  analysis  below, we  work with the minimally coupled massless and massive case ($\xi=0,\,m=\phm$), as well as the conformally coupled massless case ($\xi=\frac{(d - 2)}{4(d-1)}$, $m_p=0$). Note that the special case in $4d$ de Sitter of $m=m_c=\sqrt{2}$ is just the conformally coupled massless case.    

\section{Causal Set SJ Vacuum from Simulations} 
\label{sec:numerics} 
We now present our numerical simulations for the causal set SJ vacuum in the causal diamonds in 2d and 4d Minkowski spacetime and slabs of 2d and 4d global de Sitter spacetime. 
Where visible, error bars in the binned data reflect the SEM. 
\subsection{ Causal Diamond in 2d Minkowski Spacetime}
We begin by revisiting the analysis of $\wsj$ for the massless FSQFT in a causal diamond in 2d Minkowski spacetime  \cite{Afshordi:2012ez}. 
The IR-regulated Minkowski two-point function is
\begin{equation}
  \text{Re}[\Wmink]=-\frac{1}{2 \pi} \ln(x) + c_1,\quad \quad x=\tau \,\text{or}\, |d|, 
  \label{eq:2dmink}
\end{equation}

where $c_1$ depends on the IR cutoff. In \cite{Afshordi:2012ez} it was shown that in a small subregion in the center of the causal diamond (i.e. away from the boundaries)

\begin{equation}
c_1\approx -\frac{1}{2 \pi} \ln(\lambda e^\gamma),
\end{equation}
where $\gamma$ is the Euler-Mascheroni constant and $\lambda\sim0.46/L$, and where $2L$ is the side length of the diamond.   

In our simulations, we work in units where the volume (in 2d this is an area) of the diamond is  unity, $L=1/2,\,V=4L^2=1$. Therefore, when we compare to the continuum function \eqref{eq:2dmink}, we set $c_1\approx -0.0786$.  

\begin{figure}[H]
\begin{center} 
\includegraphics[width = .7\textwidth]{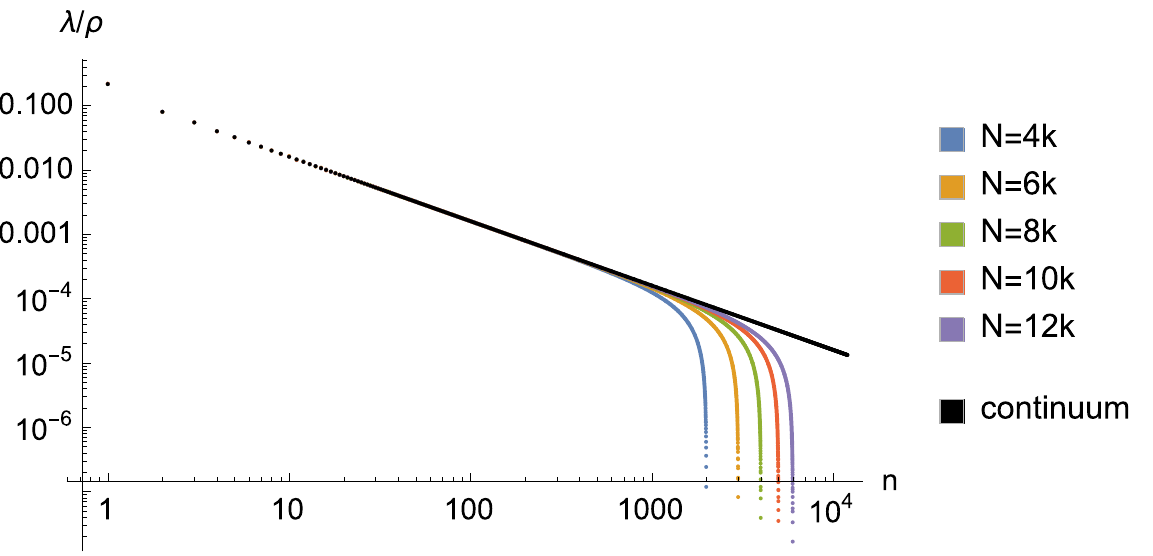}
\caption{Log-log plot of the eigenvalues of $i\Delta$ divided by density $\rho$ (except for the continuum), in the 2d causal diamond; $m=0$.}
\label{2dcdeigens}
\end{center} 
\end{figure}

\begin{figure}[H]
\begin{center} 
\subfloat[Causal]{\includegraphics[width = .46 \textwidth]{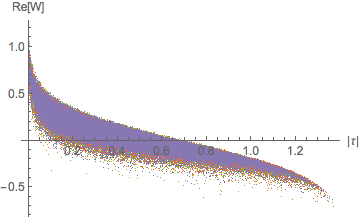}}
\subfloat[Spacelike]{\includegraphics[width = .53 \textwidth]{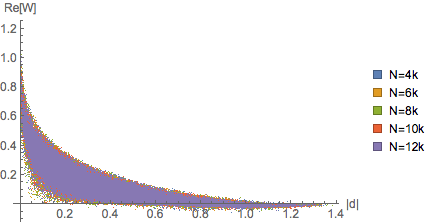}}\\
\subfloat[Causal]{\includegraphics[width = .46 \textwidth]{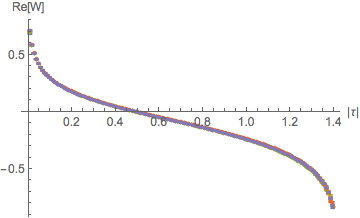}}
\subfloat[Spacelike]{\includegraphics[width = .53 \textwidth]{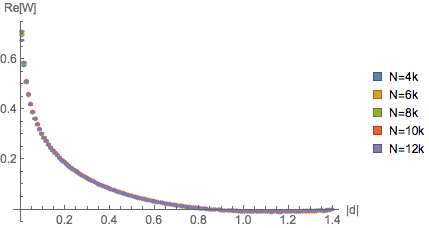}}
\caption{(a)-(b) represent $\tre[\wsj]$ vs. geodesic distance for a sample of $100000$ randomly selected pairs, in the 2d causal diamond; $m=0$. (c)-(d) are plots of the binned and averaged data with the SEM.}
\label{2dcd}
\end{center} 
\end{figure}

\begin{figure}[H]
\begin{center} 
\subfloat[Causal]{\includegraphics[width = .45 \textwidth]{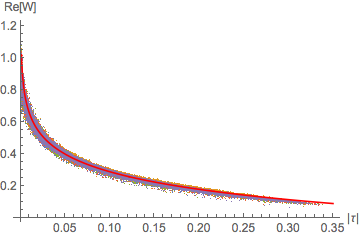}}
\subfloat[Spacelike]{\includegraphics[width = .56 \textwidth]{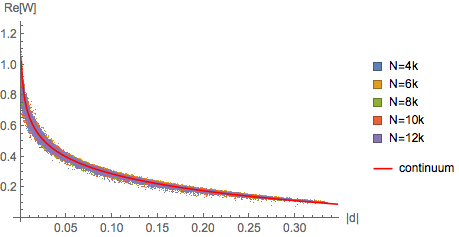}}\\
\subfloat[Causal]{\includegraphics[width = .45 \textwidth]{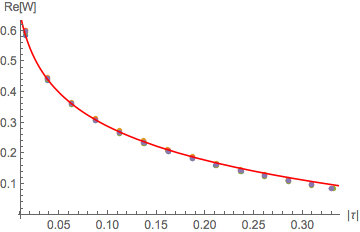}}
\subfloat[Spacelike]{\includegraphics[width = .56 \textwidth]{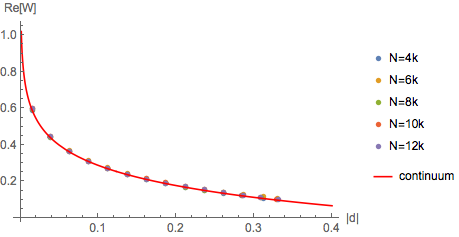}}
\caption{(a)-(b) represent $\tre[\wsj]$ vs. geodesic distance for all pairs within a sub-diamond with side length $1/4$ of that the full diamond, in the 2d causal diamond; $m=0$. (c)-(d) are plots of the binned and averaged data with the SEM. In both cases, the continuum IR-regulated Minkowski Wightman function  \eqref{eq:2dmink} has also been shown.}
\label{2dcdsub}
\end{center} 
\end{figure}

Our results are shown in figures \ref{2dcdeigens}-\ref{2dcdsub} and agree with the ab initio construction of \cite{Afshordi:2012ez}. Figure \ref{2dcdeigens} is a log-log plot of the positive causal set SJ eigenvalues, along with the positive continuum eigenvalues (discussed at the end of Section \ref{sjvac}). The two sets of eigenvalues are in agreement up to a characteristic ``knee" at which the causal set spectrum dips and ceases to obey a power-law with exponent $-1$.\footnote{This behaviour and its role in calculating the SEE are discussed in  \cite{Yazdi:2017pbo}.} There is a clear convergence of the spectrum with causal set size $N$ except that the knee is pushed to smaller eigenvalues as $N$ increases. 

Figure \ref{2dcd} shows scatter plots of $\text{Re}[\wsj]$ for pairs of events that are causally and spacelike related; it also shows the binned and averaged plots where the convergence becomes clear. The convergence with $N$ is very good and tells us that we are in the asymptotic regime. This is the kind of convergence  we will look for when either a comparison with the continuum is not possible or when there is a marked discrepancy with the continuum. In order to compare with the continuum, $\wsj$ was calculated in \cite{Afshordi:2012ez} for pairs of points in a small causal diamond in the center of the larger causal diamond and it was shown that $\wsj$ agreed with the Minkowski vacuum in \eqref{eq:2dmink}. We carry out a similar comparison and the results are shown in figure \ref{2dcdsub}. This figure shows the scatter plots and the binned and averaged plots for $\wsj$ within a smaller  diamond of side length $1/4$ compared to that of the original diamond it is concentric to. The continuum IR-regulated Minkowski curve is also plotted. These plots confirm that away from the boundaries of the diamond $\tre[\wsj]$ indeed resembles the Minkowski vacuum, as was shown analytically and numerically in \cite{Afshordi:2012ez}.  

\subsection{ Causal Diamond in 4d Minkowski Spacetime}
Next we examine the massless FSQFT in a causal diamond in 4d Minkowski spacetime. Unlike in 2d, we do not have an analytic ab initio calculation to compare with or refer to. We will instead rely on convergence properties and comparisons with the  continuum in a small causal diamond within the larger one. Another difference with the 2d case is that the causal set retarded Green function only agrees with the continuum one in the infinite density limit. This was discussed above in Section \ref{subsec:green}.
 
The 4d Minkowski two-point function is 
\begin{equation}
    \text{Re}[\Wmink]=\frac{1}{4 \pi^2 x^2},\quad x=i\tau \,\text{or}\, |d| .
    \label{eq:4dmink}
\end{equation}

\begin{figure}[H]
\begin{center} 
\includegraphics[width = .56 \textwidth]{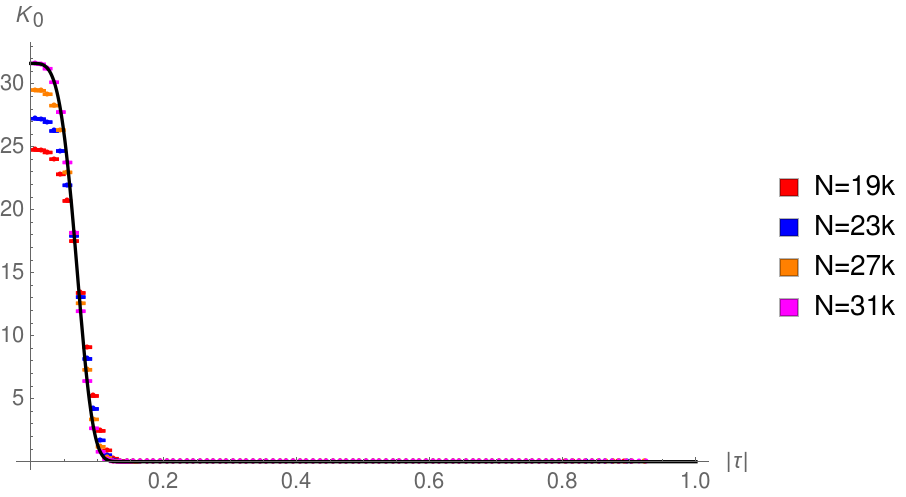}
\caption{The binned and averaged plot for $K_0$ vs. $|\tau|$ as $N$ is varied, in the 4d causal diamond. The black curve represents the expectation value  \eqref{masslessgreenexp} for $N=31k$. We see an excellent match.}
\label{4dcdgreen}
\end{center} 
\end{figure}

We work in units where the (top to bottom corner) height of the diamond is unity.
In figure \ref{4dcdgreen} we plot binned and averaged values for the causal set retarded Green  function \eqref{masslessgreen} along with its  expectation value at finite density \eqref{masslessgreenexp}. The corresponding continuum Green function \eqref{masslessgreencont} has a delta function on the lightcone and is therefore infinitely sharply peaked there. While this is not the case in the causal set, the discrepancy grows smaller as the density is increased.

\begin{figure}[H]
\begin{center} 
\includegraphics[width = .6 \textwidth]{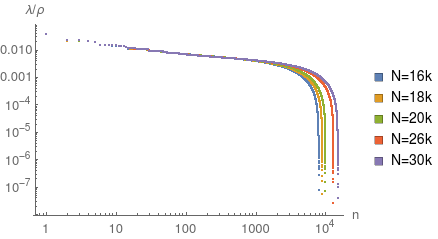}
\caption{Log-log plot of the eigenvalues of $i\Delta$ divided by density $\rho$, in the 4d causal diamond; $m=0$.}
\label{4dcdeigens}
\end{center} 
\end{figure}

In figure \ref{4dcdeigens} we show the log-log plot of the SJ spectrum. This spectrum is qualitatively similar to the spectrum in the 2d diamond, in that it obeys a power-law in the large eigenvalue regime,  while exhibiting a knee in the UV (smaller eigenvalue regime) where it dips. It moreover converges well as $N$ is increased, except near the knee which, as in the 2d diamond, shifts to the UV as $N$ increases. This suggests that we are in the asymptotic regime. 

\begin{figure}[H]
\begin{center} 
\subfloat[Causal]{\includegraphics[width = .46 \textwidth]{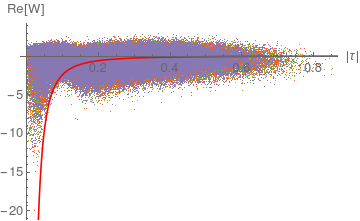}}
\subfloat[Spacelike]{\includegraphics[width = .53 \textwidth]{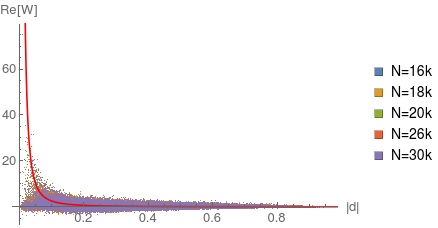}}\\
\subfloat[Causal]{\includegraphics[width = .46 \textwidth]{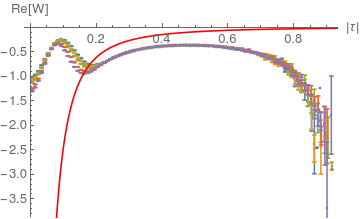}}
\subfloat[Spacelike]{\includegraphics[width = .53 \textwidth]{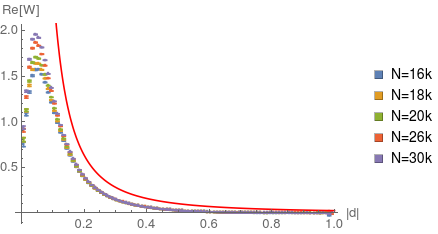}}
\caption{(a)-(b) represent $\tre[\wsj]$ vs. geodesic distance for a sample of $100000$ randomly selected pairs, in the 4d causal diamond; $m=0$. (c)-(d) are plots of the binned and averaged data with the SEM. In both cases, the continuum Minkowski Wightman function \eqref{eq:4dmink} has also been shown in red.}
\label{4dcd}
\end{center} 
\end{figure}

In figure \ref{4dcd} we show the scatter and binned plots for $\text{Re}[\wsj]$ as $N$ is varied. The convergence with increasing density suggests that the larger $N$ values are approaching the asymptotic regime. The Minkowski two-point function \eqref{eq:4dmink} is also included in this plot and it clearly does not agree with $\wsj$ in the full diamond. The small distance behaviour shows an interesting departure from the continuum, softening the divergences. 

\begin{figure}[H]
\begin{center} 
\subfloat[Causal]{\includegraphics[width = .45 \textwidth]{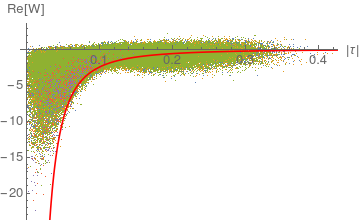}}
\subfloat[Spacelike]{\includegraphics[width = .54 \textwidth]{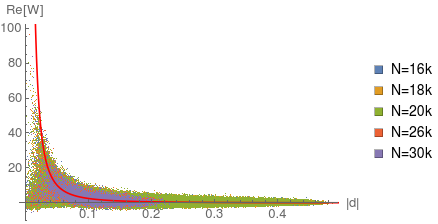}}\\
\subfloat[Causal]{\includegraphics[width = .45 \textwidth]{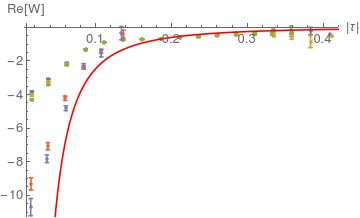}}
\subfloat[Spacelike]{\includegraphics[width = .54 \textwidth]{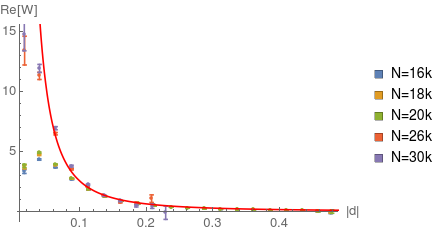}}
\caption{(a)-(b) represent $\tre[\wsj]$ vs. geodesic distance for all pairs within a sub-diamond with height $1/2$ of the full diamond, in the 4d causal diamond; $m=0$. (c)-(d) are plots of the binned and averaged data with the SEM. In both cases, the continuum Minkowski Wightman function  \eqref{eq:4dmink} has also been shown.}
\label{4dcdsub}
\subfloat[Causal pairs]{\includegraphics[width=.5\linewidth]{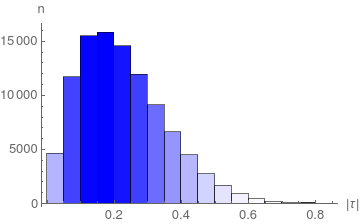}}
\subfloat[Spacelike pairs]{\includegraphics[width=.5\linewidth]{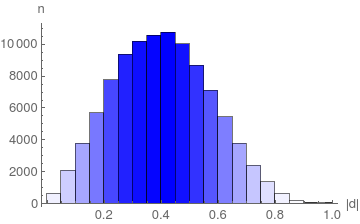}}
\caption{Distribution of the number of causal and spacelike pairs $n$ with magnitude of the geodesic distance for $N=30k$, in the 4d causal diamond.}
\label{4dcdpairs}
\end{center} 
\end{figure}

Figure \ref{4dcdsub} shows the scatter and binned plots for a smaller causal diamond of side length $1/2$ compared to the larger diamond it is in the center of. Although the agreement of $\wsj$ with $\Wmink$ is not as good as in 2d, we see that as $N$ increases, there is a convergence of $\wsj$ to $\Wmink$. This suggests that as in 2d, the 4d diamond also shows an agreement with the Minkowski vacuum far away from the boundary. 

Figure \ref{4dcdpairs} shows the distribution of pairs of points in the diamond as a function of the proper time and  distance. From this plot one can see that there are many fewer pairs of points at small and large proper distance and times than in the intermediate regimes. Nevertheless, the scatter plots and the error bars on the binned plots do not show significant deviation in these regimes. 

\subsection{Slab of 2d de Sitter Spacetime}
The simulations in the 2d and 4d causal diamond help set the stage for the simulations in  slabs of 2d and 4d de Sitter spacetime, which we turn to in this and the next subsection. As in the causal diamond examples, we will look for convergence of the causal set calculation with $N$ to establish that we are in the asymptotic regime. The slab in de Sitter spacetime lies within the region $[-T,T]$\footnote{$T$ is the cutoff in the conformal time defined in  \eqref{confmetric2}.} and we will probe our results' sensitivity to  $T$. We will also look for convergence with $T$ at fixed $\rho$, to show that the results are independent of the cutoff. 

The Wightman function for the Euclidean vacuum in $d$ spacetime dimensions is given by\footnote{The expression for $W_E$ in equation $B.36$ of \cite{Aslanbeigi:2013fga} has a minor typographical error: the factor of $4\pi$ should be raised to the power of $d/2$. See for example \cite{Bousso:2001mw}.}
\begin{equation}
W_E(x,y)=\frac{\Gamma[h_+] \Gamma[h_-]}{(4 \pi)^{d/2}\ell^2 \Gamma[\frac{d}{2}]}\,  {}_2F_1\left(h_+,h_-, \frac{d}{2};\frac{1+Z(x,y)+i\epsilon\, \text{sign}(x^0-y^0)}{2}\right),
\label{we}
\end{equation}
where $Z(x, y)$ is defined by \eqref{zdef},  $h_\pm=\frac{d-1}{2}\pm\nu$, $\nu=\ell\sqrt{m_*^2-m^2}$, $m_*=\frac{d-1}{2\ell}$ and $_2F_1(a,b,c;z)$ is a hypergeometric function. The 
symmetric two-point function, or Hadamard function, for any other Allen-Mottola $\alpha$-vacuum is \cite{Aslanbeigi:2013fga}
\begin{equation}
    H_{\alpha\beta}(x,x')=\cosh2\alpha\,H_E(x,x')+\sinh2\alpha\,[\cos\beta\,H_E(\bar{x},x')-\sin\beta\,\Delta(\bar{x},x')]
    \label{walpha},
\end{equation}
where $\bar{x}$ is the antipodal point of $x$. The Wightman function is related to $H$ by $2W=H+i\Delta$. We will make comparisons with the $\alpha$-vacua found to correspond to the SJ vacuum in \cite{Aslanbeigi:2013fga}. Since we  work in even dimensions, these are
$\alpha=0$ for $m\geq m_*$ (yielding the Euclidean vacuum), and 
\begin{equation}
    \alpha=\frac{1}{2}\tanh^{-1}|\sin\pi\nu|\quad\text{and}\quad\beta=\pi[\frac{d}{2}+\theta(-\sin\pi\nu)]
\end{equation}
for $m<m_*$.

\begin{figure}[H]
\subfloat[$m=0$]{\includegraphics[width = .43\linewidth]{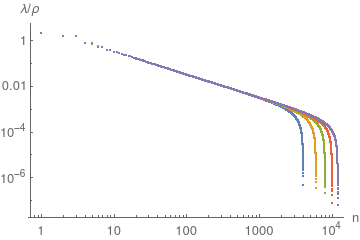}} 
\subfloat[$m=2.3$]{\includegraphics[width = .57\linewidth]{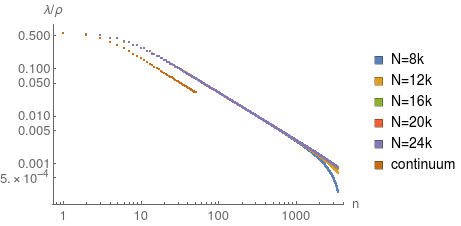}}
\caption{Log-log plot of the positive eigenvalues of $i\Delta$ at $T=1$, in 2d de Sitter. In the massive case on the right we plot the largest 3500 positive eigenvalues and the corresponding continuum eigenvalues from the finite T mode comparison results of \cite{Aslanbeigi:2013fga}.} 
\label{fig:2ddSspectrum}
\end{figure}

In this subsection we consider 2d de Sitter spacetime, and work in units in which the de Sitter radius $\ell=1$. In 2d,  $m_*= 0.5$, and the conformal mass $m_c=0$. Hence the minimally coupled and the conformally coupled massless cases coincide.
Our simulations span slabs of different heights given by $T$ values ranging from $1$ to $1.5$, while our $N$ values range from $8k$ to $36k$. We show the log-log plots of the PJ spectrum for the massless $m=0$ and for the massive $m=2.3$ cases in figure \ref{fig:2ddSspectrum}. As in the 2d diamond, the causal set spectrum exhibits a characteristic knee. The spectrum converges very well for both sets of masses, with the knee shifting to the UV as $N$ increases, as expected. We also compare the causal set spectrum with the finite $T$ continuum spectrum obtained via the mode comparison method in \cite{Aslanbeigi:2013fga}. As shown in figure \ref{fig:2ddSspectrum} this spectrum does not seem to agree with the causal set spectrum even though the latter convergences with $N$. 

\begin{figure}[H]
\begin{center}
 \subfloat[Causal $m=2.3$]{\includegraphics[width=.45\linewidth]{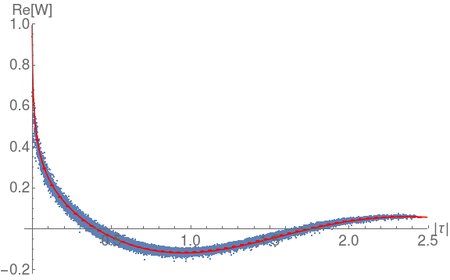}}
\subfloat[Spacelike $m=2.3$]{\includegraphics[width=.45\linewidth]{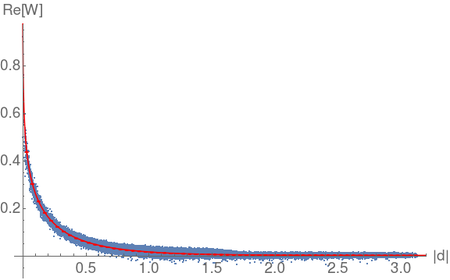}}
\caption{$N=24000,\, T=1,\, \rho=1226.31$, in 2d de Sitter. The scatter plot is $\tre[\wsj]$ vs. geodesic distance for a sample of 100000 randomly selected
pairs. The red curve represents the continuum $W_E$ from \eqref{we}.}
\label{2ddSSJ1}
\end{center} 
\end{figure}

\begin{figure}[H]
\begin{center} 
 \subfloat[Causal $m=2.3$]{\includegraphics[width=.45\linewidth]{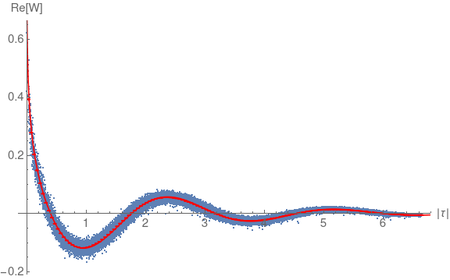}}
  \subfloat[Spacelike $m=2.3$]{\includegraphics[width=.45\linewidth]{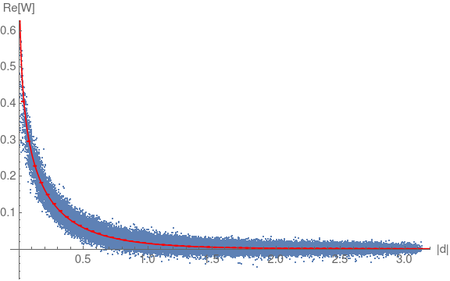}}
  \caption{$N=36000,\, T=1.5,\, \rho=203.15$, in 2d de Sitter.  $\tre[\wsj]$ vs. geodesic distance for 100000 randomly selected
pairs. The red curve represents the continuum $W_E$ from \eqref{we}.}
  \label{2ddSSJ15}
\end{center} 
 \end{figure}

\begin{figure}[H]
\begin{center}
\subfloat[Causal $m=2.3$. ]{\includegraphics[width=.45\linewidth]{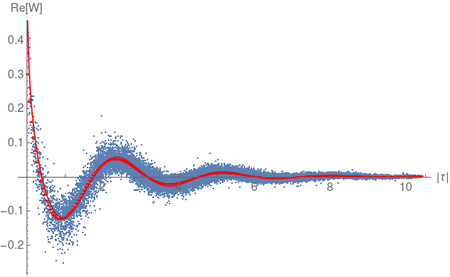}}
\subfloat[Spacelike  $m=2.3$]{\includegraphics[width = .45\linewidth]{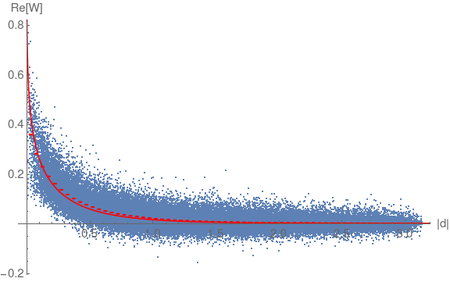}}
\caption{$N=36000, T=1.56, \rho=30.93$, in 2d de Sitter. $\tre[\wsj]$ vs. geodesic distance for a sample of 100000 randomly selected pairs. The red curve represents the continuum $W_E$ from \eqref{we}.}
\label{2ddST156}
\end{center} 
 \end{figure}

In the simulations whose results we present below, we examine two masses in detail: $m=0$ and $m=2.3$\footnote{This is an arbitrary choice of mass with no special physical significance. It  allows for comparisons with \cite{Aslanbeigi:2013fga} who use a similar mass in their 2d de Sitter causal set simulations.}, and vary over both the slab height $T$ as well as the density $\rho$. For $m=2.3$, as can be seen in the scatter plots of figures \ref{2ddSSJ1}, \ref{2ddSSJ15} and \ref{2ddST156}, $\wsj$ agrees very well with the SJ vacuum expected from the calculation in \cite{Aslanbeigi:2013fga} (the Euclidean vacuum). Furthermore, it appears that $\wsj$ for a given $T$ is simply the restriction of $\wsj$ for a larger $T$. This is also in agreement with the simulation results of \cite{Aslanbeigi:2013fga}. 

\begin{figure}[H]
\begin{center}   
 \subfloat[Causal $m=0$]{\includegraphics[width=.42\linewidth]{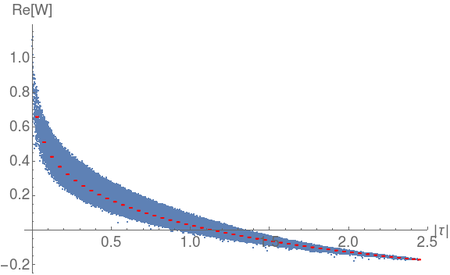}}
\subfloat[Spacelike $m=0$]{\includegraphics[width=.42\linewidth]{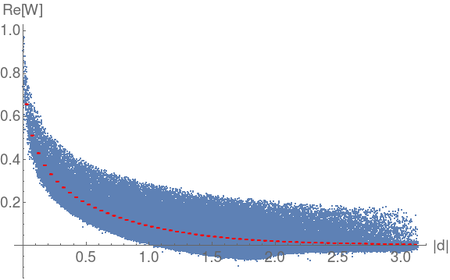}}\\
\subfloat[Causal pairs]{\includegraphics[width=.4\linewidth]{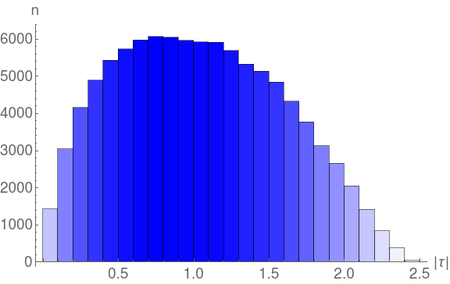} }
\subfloat[Spacelike pairs]{\includegraphics[width=.4\linewidth]{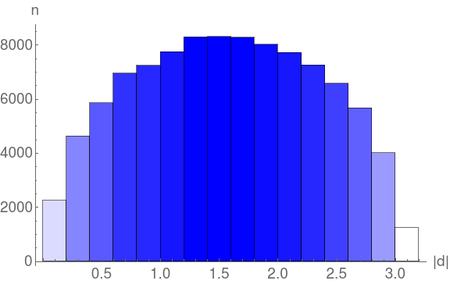}}
\caption{$N=32000, T=1, \rho=1635.08$, in 2d de Sitter. (a)-(b) represent $\tre[\wsj]$ vs. geodesic distance for a sample of 100000 randomly selected
pairs, and the red curve represents the mean values with the SEM. (c)-(d) are plots of the distribution of pairs.}
\label{2ddSSJ1zero}
\end{center}
\end{figure}

\begin{figure}[H]
\begin{center} 
 \subfloat[Causal $m=0$]{\includegraphics[width=.45\linewidth]{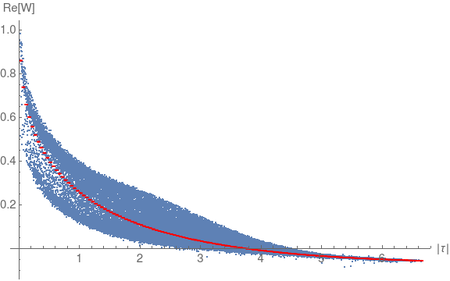}}
  \subfloat[Spacelike $m=0$]{\includegraphics[width=.45\linewidth]{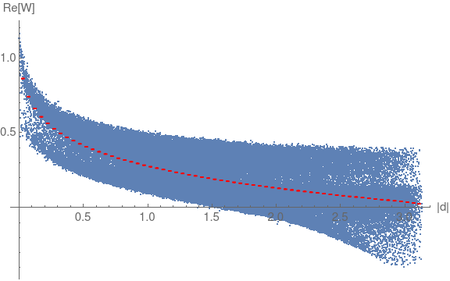}}\\
  \subfloat[Spacelike pairs]{\includegraphics[width=.45\linewidth]{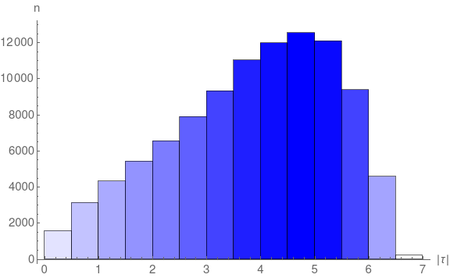}}
\subfloat[Spacelike pairs]{\includegraphics[width=.45\linewidth]{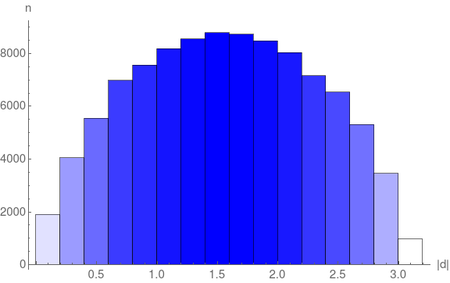}}
\caption{$N=36000, T=1.5, \rho=203.15$, in 2d de Sitter. (a)-(b) represent $\tre[\wsj]$ vs. geodesic distance for a sample of 100000 randomly selected pairs. The red curve represents the mean values with the SEM. (c)-(d) are plots of the distribution of pairs.}
\label{2ddSSJ15zero}
  \end{center} 
\end{figure}

\begin{figure}[H]
\begin{center} 
\subfloat[Causal $m=0$]{\includegraphics[width = .45\linewidth]{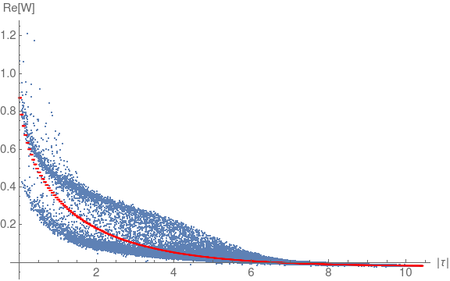}}
\subfloat[Spacelike $m=0$]{\includegraphics[width = .45\linewidth]{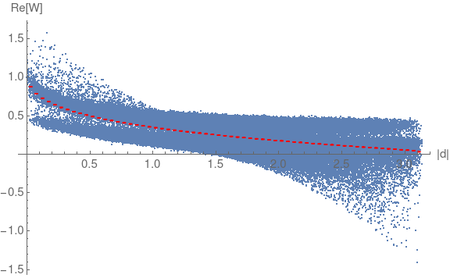}}\\
\subfloat[Causal pairs]{\includegraphics[width=.45\linewidth]{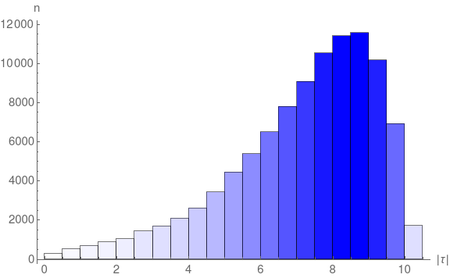}}
\subfloat[Spacelike pairs]{\includegraphics[width=.45\linewidth]{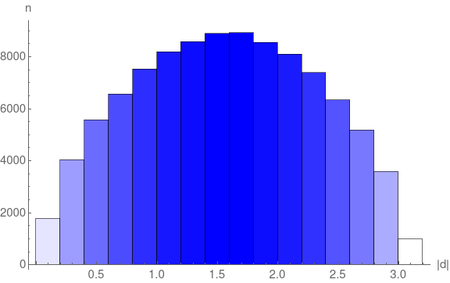}}
\caption{$N=36000, -1.56<\tilde{T}<1.56, \rho=30.93$, in 2d de Sitter. (a)-(b) represent $\tre[\wsj]$ vs. geodesic distance for a sample of  100000 randomly selected pairs. The red curve represents the mean values (of the data) with the SEM. (c)-(d) are plots of the distribution of pairs.}
  \label{2ddST156zero}
\end{center} 
 \end{figure}

For the massless case, the scatter plots of $\wsj$ in figures \ref{2ddSSJ1zero}, \ref{2ddSSJ15zero} and \ref{2ddST156zero} do not show convergence, but instead fan out, as a function of the proper time and distance. As the density decreases, for $T=1.56, N=36k$, the scatter plot figure \ref{2ddST156zero} shows a clustering into two distinct sets. This shows that $\wsj$ may not just be a function of proper time and distance, and hence {\it may not be} de Sitter invariant.

\begin{figure}[H]
      \begin{center}
          \subfloat[Causal $m=0$]{\includegraphics[width=.46\textwidth]{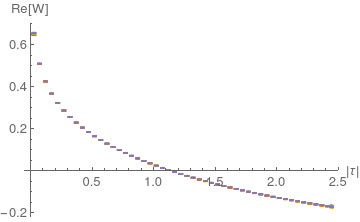}}
           \subfloat[Spacelike  $m=0$]{\includegraphics[width=.53\textwidth]{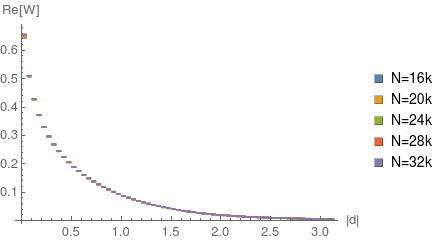}}\\
            \subfloat[Causal $m=2.3$]{\includegraphics[width=.46\textwidth]{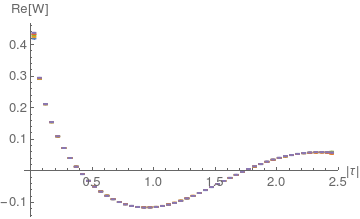}}
             \subfloat[Spacelike  $m=2.3$]{\includegraphics[width=.53\textwidth]{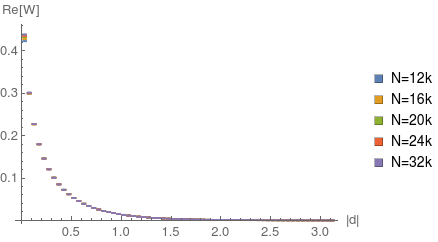}}
      \end{center}
      \caption{Variation of binned and averaged $\tre[\wsj]$ with density at $T=1$, in 2d de Sitter.}
      \label{fig:2ddSsem}
  \end{figure}

In figure \ref{fig:2ddSsem} the binned and averaged plots for $\wsj$ show  very good convergence with $N$. While this is consistent with the narrowing of the $m=2.3$ scatter plots at higher densities, the convergence for $m=0$ is not (since the $m=0$ scatter plots do not narrow much). Hence both the scatter plots and the binned plots are important in determining convergence as well as understanding the nature of $W_{SJ}$. 

 
\subsection{Slab of 4d de Sitter Spacetime}

Finally, we examine the 4d de Sitter SJ vacuum. Again, we work in units in which the de Sitter radius $\ell=1$. In 4d, $m_*=1.5$ and $m_c=\sqrt{2}\approx 1.41$. 

\begin{figure}[H]
\begin{center}
\includegraphics[width=.52\linewidth]{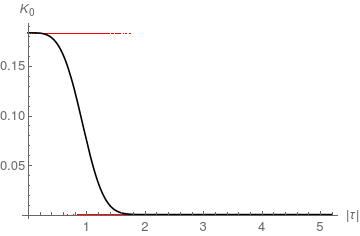}
\caption{$K_0$ vs. $|\tau|$ for $N=32k, T=1.42,\rho=7.978,m=m_c=\sqrt{2}$, in 4d de Sitter. The black curve represents the expectation value \eqref{masslessgreenexp}.}
\label{fig:4ddSgreen}
\end{center} 
\end{figure}

\begin{figure}[H]
   \subfloat[$m=0,\,T=1.5$]{\includegraphics[width=.46\linewidth]{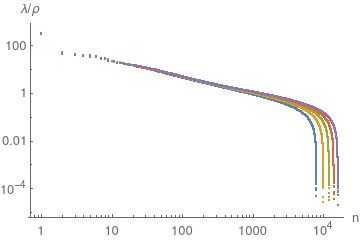}}
   \subfloat[$m=2.3,\,T=1.2$]{\includegraphics[width=.53\linewidth]{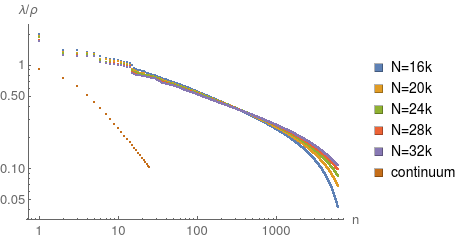}}
    \caption{Log-log plot of the positive eigenvalues of $i\Delta$, in 4d de Sitter. In the massive case on the right we plot the largest 6000 positive eigenvalues and the corresponding continuum eigenvalues from the finite T mode comparison results of \cite{Aslanbeigi:2013fga}.}
    \label{fig:4ddSeigen}
\end{figure}

\begin{figure}[H]
\begin{center}
   \subfloat[Near $m=0$]{\includegraphics[width=.47\linewidth]{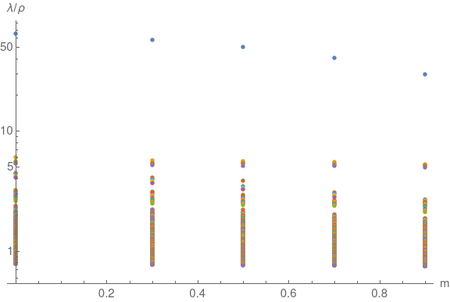}}
   \subfloat[Near $m=1.41$]{\includegraphics[width=.47\linewidth]{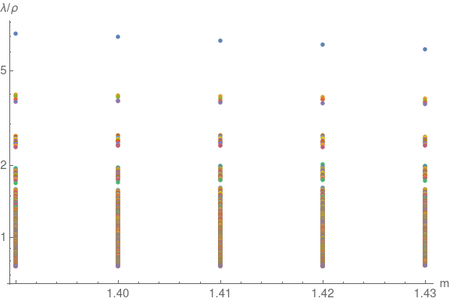}}
    \caption{Log-linear plot of the first 500 positive eigenvalues of $i\Delta$ at $T=1.42,\, \rho=9$, in 4d de Sitter.}
    \label{fig:4ddSeigenwithm}
\end{center}
\end{figure}

In figure \ref{fig:4ddSgreen} we show the scatter plot of the causal set retarded Green function \eqref{retgreendS}, taking the conformally coupled massless case as an example. While the small $\tau$ discrepancy with  the continuum expression is expected and attributed to the local finiteness of the causal set, the behaviour for large $\tau$ compares well with the continuum.
Figure \ref{fig:4ddSeigen} shows the log-log plot of the SJ spectrum for $m=0$ and $m=2.3$ for various $N$. We find that there is excellent convergence with $N$ in both cases, and again, as in the other cases we have seen, there is a knee which shifts to the UV as $N$ is increased. However, there is poor agreement with the continuum values of the finite $T$ spectrum calculated via the mode comparison method in \cite{Aslanbeigi:2013fga}, as in the 2d case. In figure \ref{fig:4ddSeigenwithm} we also show the spectrum for $m$ varied around $m=0$ and $m=m_c\approx1.41$. There is no unusual behaviour close to these masses.    

\begin{figure}[H]
\subfloat[Causal $T=1$]{\includegraphics[width=.48\linewidth]{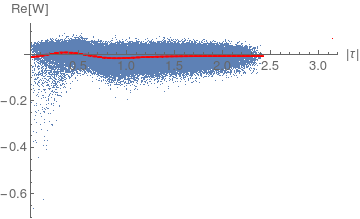}}
\subfloat[Spacelike $T=1$]{\includegraphics[width=.48\linewidth]{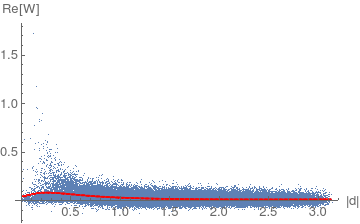}}\\
\subfloat[Causal $T=1.2$] {\includegraphics[width=.48\linewidth]{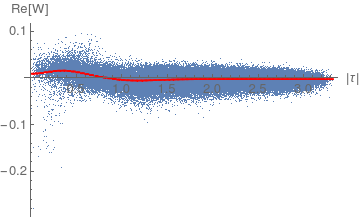}}
\subfloat[Spacelike $T=1.2$]{\includegraphics[width=.48\linewidth]{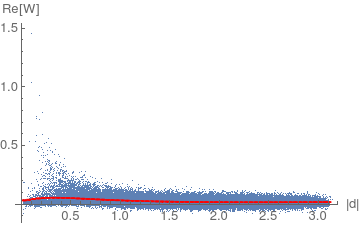}}
\caption{$m=0,\,N=32000$, in 4d de Sitter. $\tre[\wsj]$ vs. geodesic distance for   100000 randomly selected pairs,  and the red curve represents the mean
values with the SEM.}
\label{fig:4ddSzero}
\end{figure}

\begin{figure}[H]
  \begin{center}
\subfloat[Causal $T=1$]{\includegraphics[width=.45\linewidth]{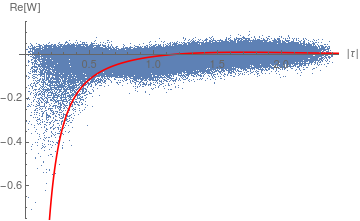}}
\subfloat[Spacelike $T=1$]{\includegraphics[width=.45\linewidth]{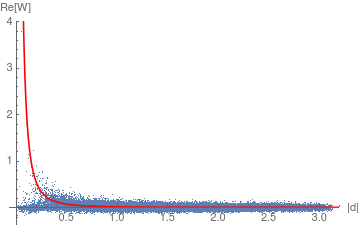}}\\
\subfloat[Causal $T=1.2$]{\includegraphics[width=.5\linewidth]{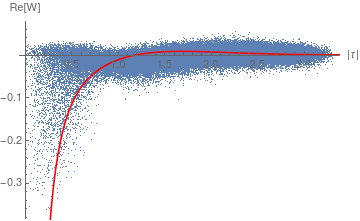}}
\subfloat[Spacelike $T=1.2$]{\includegraphics[width=.5\linewidth]{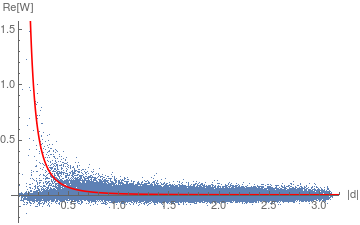}}
\caption{$m=2.3,\,N=32000$, in 4d de Sitter. $\tre[\wsj]$ vs. geodesic distance for a sample of  100000 randomly selected pairs. The red curve shows the Euclidean two-point function $W_E$ from \eqref{we}.} 
\label{fig:4ddS}
  \end{center}
  \end{figure}

\begin{figure}[H]
  \begin{center}
      \subfloat[Causal $m=0,T=1.3$]{\includegraphics[width=.46\textwidth]{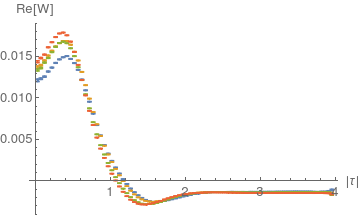}}
       \subfloat[Spacelike $m=0,T=1.3$]{\includegraphics[width=.53\textwidth]{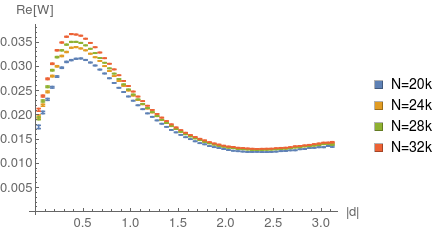}}\\
        \subfloat[Causal $m=1.41,T=1.4$]{\includegraphics[width=.46\textwidth]{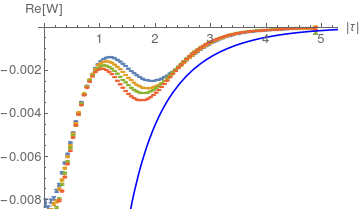}}
         \subfloat[Spacelike $m=1.41,T=1.4$]{\includegraphics[width=.53\textwidth]{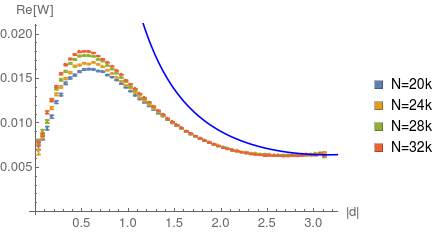}}
         \caption{$\tre[\wsj]$ vs. geodesic distance with varying density, in 4d de Sitter. The  blue curve shows the Euclidean two-point function as a reference.}
  \label{fig:4ddSfixedT}
  \end{center}
\end{figure}

\begin{figure}[H] 
\begin{center} 
\subfloat[Causal $m=0.7$]{\includegraphics[width=.46\textwidth]{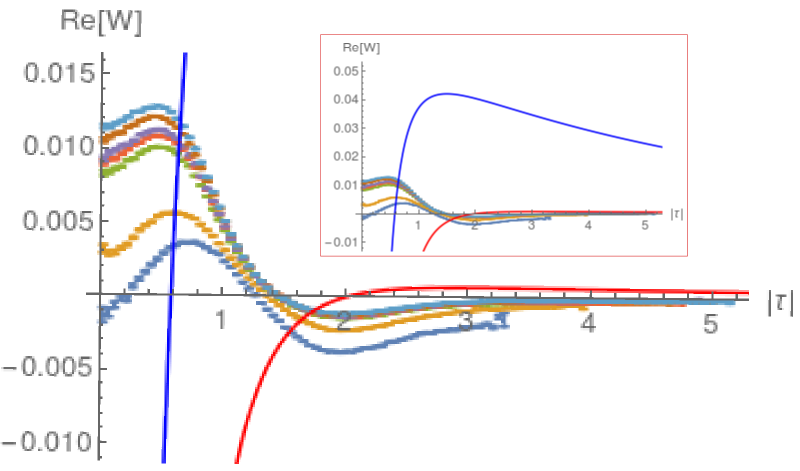}}
\subfloat[Spacelike $m=0.7$]{\includegraphics[width=.53\textwidth]{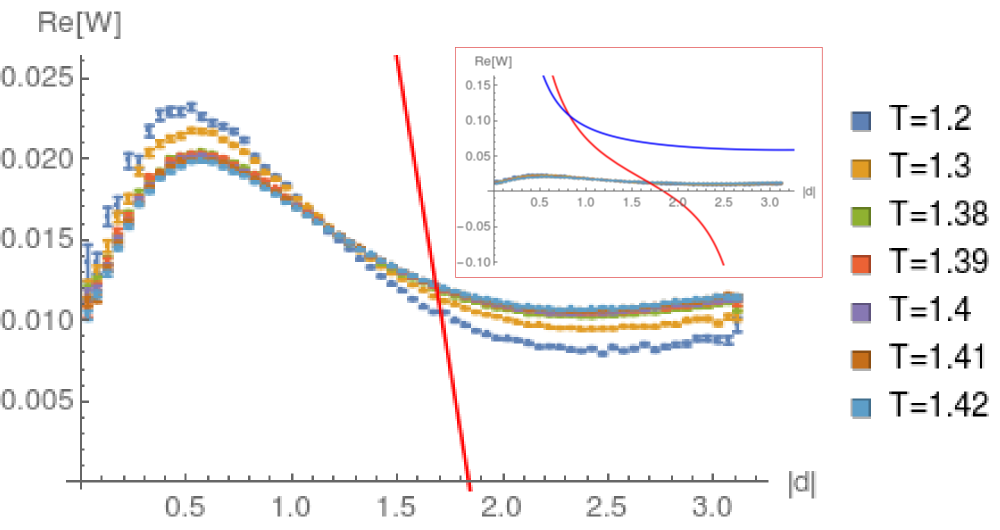}}\\
\subfloat[Causal $m=1.41$]{\includegraphics[width=.46\textwidth]{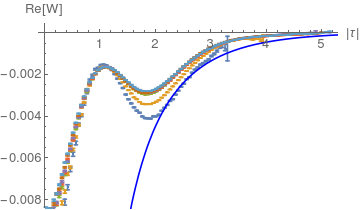}}
\subfloat[Spacelike $m=1.41$]{\includegraphics[width=.53\textwidth]{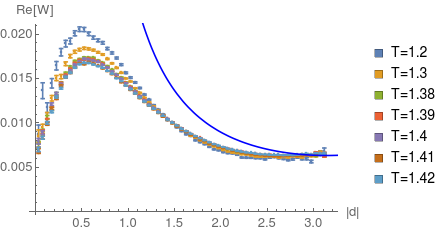}}\\
\subfloat[Causal $m=1.5$]{\includegraphics[width=.46\textwidth]{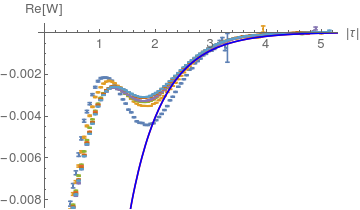}}
\subfloat[Spacelike $m=1.5$]{\includegraphics[width=.53\textwidth]{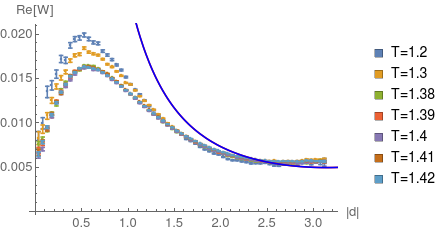}}
\caption{$\tre[\wsj]$ vs. geodesic distance with varying $T$ for various $m$ at $\rho=9$, in 4d de Sitter. The red and blue curves represent the corresponding continuum $\alpha$- and Euclidean two-point functions respectively. The inset figures represent the zoomed-out versions. In (e)-(f), for $m=\sqrt{2}$ there is no corresponding $\alpha$-vacuum, and in (g)-(h) the $\alpha$-vacuum and Euclidean vacuum coincide.}
\label{4ddSfixedrho}
\end{center} 
\end{figure}

Figures \ref{fig:4ddSzero} and \ref{fig:4ddS} are sample scatter plots of $\wsj$ for $m=0$ and $m=2.3$.
In figure \ref{fig:4ddSfixedT} we fix $T$ for $m=0$ and for $m=m_c\approx1.41$ and vary $N$ to check for convergence with density; for smaller proper times and distances, the convergence is not as good as it is for larger proper times and distances. For $m=1.41$ we also plot the  Wightman function associated with the  Euclidean vacuum $W_E$ in \eqref{we}. $W_E$ does not compare well with the causal set $\wsj$.
Next, in figure \ref{4ddSfixedrho} we fix the density $\rho=9$ and check the convergence with $T$, which we vary from $1.2$ to $1.42$. We find good convergence for various $m$ values. However, the Wightman function associated with the  $\alpha$-vacuum \eqref{walpha} as well as the Euclidean vacuum $W_E$ once again do not compare well with the causal set $\wsj$ for any of these masses. This is somewhat surprising, since the discrepancy occurs well away from the massless minimally and conformally coupled cases.  

\begin{figure}[!ht]
\begin{center}
\setcounter{subfigure}{0}
\subfloat[Causal $T=1.42$]{\includegraphics[width=.45\textwidth]{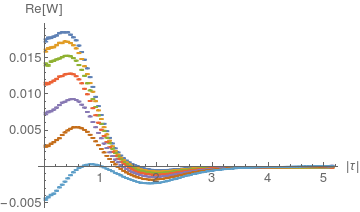}}
\subfloat[Spacelike $T=1.42$]{\includegraphics[width=.54\textwidth]{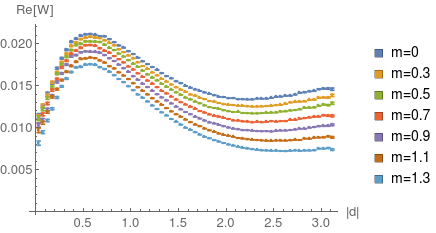}}\\
\subfloat[Causal $T=1.42$]{\includegraphics[width=.46\textwidth]{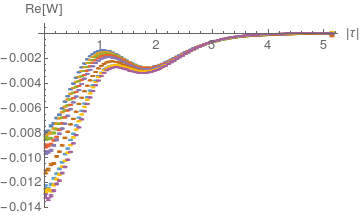}}
\subfloat[Spacelike $T=1.42$]{\includegraphics[width=.53\textwidth]{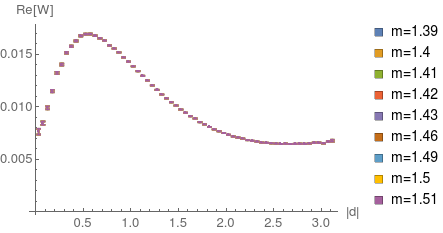}}\\
\subfloat[Causal pairs]{\includegraphics[width=.5\linewidth]{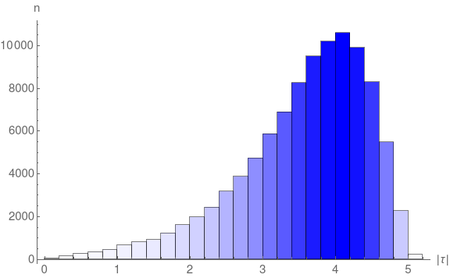}}
\subfloat[Spacelike pairs]{\includegraphics[width=.5\linewidth]{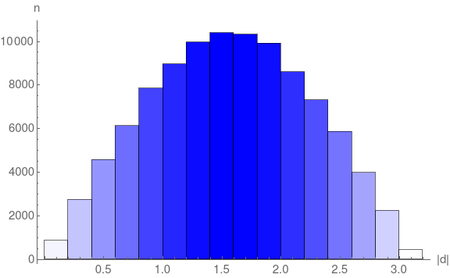}}
\caption{$\tre[\wsj]$ vs. geodesic distance with varying $m$ at $\rho=9$, in 4d de Sitter. (e)-(f) show the distribution of pairs.}
\label{4ddSfixedrhom}
\end{center}
\end{figure}

Further, in figure \ref{4ddSfixedrhom} we look at $\wsj$ for varying masses at fixed $T=1.42$ and $\rho=9$. We find that $\wsj$ looks like a continuous function of $m$ even as $m$ is varied around $m_c$. Indeed, the  large distance behaviour for all the masses is exactly the same. At smaller distances, there is an interesting bifurcation as $m$ changes: $\text{Re}[W]$ is positive for small masses and negative for large masses.
This figure also shows the number of pairs as a function of distances. The discrepancies in the small distance behavior could be attributed to the small number of pairs there.

Our simulations thus strongly suggest that the causal set 4d de Sitter $\wsj$  differs from the Mottola-Allen $\alpha$-vacua for all masses. 
\begin{figure}[H] 
\begin{center} 
\subfloat[causal $m=0$]{\includegraphics[width=.46\textwidth]{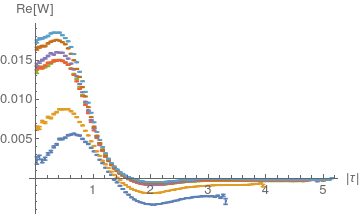}}
\subfloat[spacelike $m=0$]{\includegraphics[width=.53\textwidth]{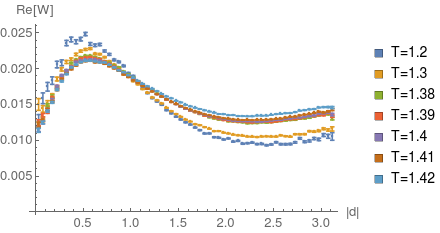}}
  \end{center}
\end{figure}
\section{Discussion}
\label{sec:discussion} 
Our simulations suggest that the CST 4d de Sitter SJ vacuum for all masses, while de Sitter invariant, is not equivalent to any of the Mottola-Allen $\alpha$-vacua. Moreover, contrary to the conclusions of \cite{Aslanbeigi:2013fga} which are based on a mode comparison calculation, we find that the CST SJ vacuum is well-defined both for $m=0$ and $m=m_{c}$ in 2d and 4d de Sitter. In 2d, where these two masses are equal, the CST SJ vacuum does not seem to be de Sitter invariant. In 4d on the other hand, as already mentioned above, the massless (as well as $m=m_{c}$) de Sitter CST SJ vacuum is de Sitter invariant. 

Our simulations are by necessity limited to a finite region of de Sitter, given by the IR cutoff $T$ and a finite density $\rho$. However, the convergence results we find are convincing and indicate that the CST SJ vacua will not change significantly as $T  \rightarrow \pi/2$ (the infinite volume limit). The convergence with density is especially good at larger proper times and distances. At smaller proper times and distances there is an approach to an asymptotic form, though not exact convergence. Put together these results suggest that the CST SJ vacuum converges to a continuum SJ vacuum with the two-point function approximately given by figures \ref{fig:4ddSfixedT} in Section \ref{sec:numerics}.    

Our results show a discrepancy with the results of \cite{Aslanbeigi:2013fga} in 4d de Sitter spacetime. One possibility, as with any numerical finding, is simply that our densities and $T$ values are not large enough to make the comparison. However, it seems an unlikely explanation given the apparent convergence we have found with density and $T$. We believe that it instead arises from the differences in how IR limits enter into the ab initio versus the mode comparison calculations. Thus, our work strongly suggests that the SJ state for 4d de Sitter is an altogether new de Sitter vacuum. 

The SJ vacuum in de Sitter spacetime clearly requires further study. An analytic ab initio calculation in the continuum is challenging, but perhaps can be carried out in a corner of the parameter space. Moreover, since the SJ state is the unique state that satisfies  \eqref{eq:sjdefn}, each of the Mottola-Allen $\alpha$-vacua must violate at least one of the SJ conditions. These ideas are currently being investigated. 
From the CST perspective, our results bring new light to questions of relevance to early universe phenomenology. Given that the continuum is an approximation to an underlying causal set, the natural vacuum for FSQFT on a 4d de Sitter-like causal set {\it is} the SJ vacuum we have obtained. Since this CST SJ vacuum differs markedly from the standard continuum 4d de Sitter vacua, it suggests that early universe phenomenology could be very different from what one expects from standard continuum calculations.

\bf Acknowledgements: \rm This research was supported in part by Perimeter Institute for Theoretical Physics. Research at Perimeter Institute is supported by the Government of Canada through the Department of Innovation, Science and Economic Development Canada and by the Province of Ontario through the Ministry of Research, Innovation and Science. SS was supported by a FQXi grant FQXi-MGA-1510 and an Emmy Noether Fellowship, at the Perimeter Institute. YY acknowledges support from the Avadh Bhatia Fellowship at the University of Alberta. NX would like to thank the  Perimeter Institute for its hospitality and Abhishek Mathur for useful discussions.

\appendix

\section*{Appendix A: de Sitter Spacetime}
\label{appA}
In this appendix we review de Sitter spacetime, mostly following the discussion in  \cite{Spradlin:2001pw}. We define the coordinate systems that we work with, as well as the definition of geodesic distance that we use to evaluate or proper times and distances. 

de Sitter spacetime  $dS_d$ can be thought of as a surface in $\mathbb{M}^{d+1}$. This surface is characterized by the constraint
\begin{equation}
-X_0^2+X_1^2+...+X_d^2=\eta_{AB}X^AX^B=\frac{1}{H^2},
\label{hypecoords}
\end{equation}
where $A$ and $B$ run from $0$ to $d$. This is a hyperboloid in $\mathbb{M}^{d+1}$ with ``radius" $\ell\equiv\frac{1}{H}$. This is also, topologically, $\mathbb{R}\times S^{d-1}$, where the $S^{d-1}$ corresponds to a surface with constant $X_0$. This $(d-1)$-sphere has a radius $\equiv\frac{1}{H^2}+X_0^2$. \\
Assume that on the surface $dS_d$ we can assign coordinates $x^a$, then corresponding to each point on the surface we can define vectors $X^A(x)$, in $\mathbb{M}^{d+1}$. Each of these must satisfy (1). We can define another useful quantity as follows: 
\begin{equation}
Z(x,y)=H^2\eta_{AB}X^A(x)X^B(y)=\cos\theta.
\label{zdef}
\end{equation}
We can think of this as an inner product between two $d+1$-vectors that represent points $x$ and $y$ on the surface $dS_d$. If there is some angle $\theta$ between these two vectors in $\mathbb{M}^{d+1}$, then the above expression can be written (in exact analogy with the usual ``dot product") in terms of this angle, and the magnitude cancels out with the $H^2$ in front. \\
Now for two points on the surface separated by an angle $\theta$, the geodesic distance (in exact analogy with a sphere) is given by $d(x,y)=\frac{1}{H}\theta$, where $\frac{1}{H}$ plays the role of radius. Therefore we have \cite{Allen:1985ux}
\begin{equation}
d(x,y)=\frac{1}{H}\cos^{-1}Z(x,y).
\end{equation}
The advantage of this relation is that in general the geodesic distance is given by 
\begin{equation}
d(x,y)=\int_y^xd\mu\sqrt{\eta_{AB}\frac{dX^A}{d\mu}\frac{dX^B}{d\mu}},
\end{equation}
where $X^A(\mu)$ is a parameterized geodesic between points $x$ and $y$. In general, this integral can be difficult to evaluate. However the closed-form expression of  $Z(x,y)$ allows it to be trivially evaluated once coordinates are assigned to the surface $dS_d$. The values $Z > 1$, $Z = 1$ and $-1 < Z < 1$ correspond to pairs of points that can be joined by timelike, null, and spacelike geodesics, respectively.

A useful set of coordinates to characterize global de Sitter spacetime are the hyperbolic coordinates. In these, the metric takes the form 
\begin{equation}
ds^2= -d\tau^2+\frac{1}{H^2}\cosh^2(H\tau)\,d\Omega_{d-1}^2  ,
\label{globmetric}
\end{equation}
where $-\infty<\tau<\infty$ and $\Omega_{d-1}$ are coordinates on $S^{d-1}$. These coordinates are related to those in \eqref{hypecoords} by
\begin{eqnarray}
    X^0&=&\frac{1}{H}\sinh\tau\\\nonumber
    X^i&=&\frac{1}{H} w^i\cosh\tau,\indent i=1,..., d,
    \label{globalcoords}
\end{eqnarray}
where $w^i$ are coordinates on the sphere $S^{d-1}$:
\begin{eqnarray}
    w^1&=&\cos\theta_1,\\\nonumber
    w^2&=&\sin\theta_1\cos\theta_2,\\\nonumber
    &\vdots&\\\nonumber
    w^{d-1}&=&\sin\theta_1...\sin\theta_{d-2}\cos\theta_{d-1},\\\nonumber
    w^d&=&\sin\theta_1...\sin\theta_{d-2}\sin\theta_{d-1},
\end{eqnarray}
and where $0\leq\theta_i<\pi$ for $1\leq i\leq d-2$ and $0\leq\theta_{d-1}<2\pi$. $\sum^d_{i=1}\left(w^i\right)^2=1$ and 

\begin{equation}
    d\Omega_{d-1}^2=\sum^d_{i=1}\left(dw^i\right)^2=d\theta_1^2+\sin^2\theta_1d\theta_2^2+...+\sin^2\theta_1...\sin^2\theta_{d-2}d\theta^2_{d-1}
\end{equation}
is the metric on $S^{d-1}$.

Another useful set of coordinates are the conformal/cylindrical coordinates obtained by setting $H\,d\tau/\cosh H\tau=d\tilde{T}$ in the above metric
\begin{equation}
ds^2=\frac{1}{H^2\,\cos^2\tilde{T}}\left(-d\tilde{T}^2+d\Omega_{d-1}^2\right),
\label{confmetric2}
\end{equation}
where $-\pi/2<\tilde{T}<\pi/2$. In these coordinates the volume of a region of height $2T$ (i.e. conformal time $\tilde{T}\in[-T,T]$) and radius $\ell$ is given by
\begin{equation}
V(T,d)=\frac{2 \pi ^{d/2}\ell^d}{\Gamma(\frac{d}{2})} \int_{-T}^{T} \sec ^d\tilde{T} \, d\tilde{T} .
\end{equation}
In our cases of interest,
\begin{eqnarray}
V(T,2)&=&4\pi \ell^2\tan T\\
V(T,4)&=&\frac{4}{3}\pi^2 \ell^4\,\tan T(\cos 2T+2)\sec ^2T.
\end{eqnarray}
The following are some other useful identities relevant to de Sitter spacetime that relate the Ricci scalar $R$ to other commonly used scales -- the cosmological constant ($\Lambda$), the de Sitter radius ($\ell$) and the Hubble constant ($H$):
\begin{equation}
R=\frac{2d}{d-2}\Lambda=d\,(d-1)H^2=\frac{d\,(d-1)}{\ell^2},
\end{equation}   
\begin{equation}
\text{  where  }\quad\Lambda=\frac{(d-1)(d-2)}{2}H^2.
\end{equation}
The critical mass \footnote{For more details see \cite{Hartong:2004rra}.} is
\begin{equation}
m_*=\frac{d-1}{2\ell}.
\end{equation}  
In $d=4$, $R=4\Lambda=12H^2=12/\ell^2$ and $m_*=\dfrac{3}{2\ell}$.

\section*{Appendix B: Mode Comparison to $O(4)$ Modes}
\label{appB}
In this appendix, we evaluate the expressions that yield the Bogoliubov transformation between the SJ modes and the $O(4)$ modes, as discussed near the end of Section \ref{sec:dS}. We remind the reader that if $|r_k|=1$, then $\alpha_k$ and therefore the Bogoliubov coefficients  diverge and the transformation becomes ill-defined.
\subsubsection*{Evaluation of $r_0$ in the O(4) Case}
We put in the values of $A_0,B_0$ and substitute $\eta-\pi/2=x$, then
\begin{eqnarray}
T_0&=&\frac{2\alpha^2}{H^2}\int_0^{b'}\frac{dx}{\cos^4x}\bigg\{\bigg(x+\frac{\sin2x}{2}\bigg)^2+t\bigg\},\\
D_0&=&\frac{-2\alpha^2}{H^2}\int_0^{b'}\frac{dx}{\cos^4x}\bigg\{\bigg(x+\frac{\sin2x}{2}\bigg)^2+d\bigg\},
\end{eqnarray}
where $t=\dfrac{1}{\alpha^4}\bigg(\dfrac{1}{16}+\beta^2\bigg)$ and $d=\dfrac{-1}{\alpha^4}\bigg(\dfrac{1}{4}+i\beta\bigg)^2$. So we have 
\begin{eqnarray}
r_0=\frac{D_0}{T_0}=-\frac{\epsilon+d}{\epsilon+t}\quad\text{where}\quad\epsilon=\frac{\displaystyle\int_0^{b'}\dfrac{dx}{\cos^4x}\bigg(x+\dfrac{\sin2x}{2}\bigg)^2 }{\displaystyle\int_0^{b'}\dfrac{dx}{\cos^4x}}\,.
\end{eqnarray}
These integrals are well-behaved at the lower limit and diverge as $b'\rightarrow\pi/2$, so we can approximate them by their values near the upper limit. We get
$$\lim_{b'\rightarrow\pi/2}\epsilon=\frac{\pi^2}{4},$$
\begin{equation}
r_0=-\frac{\pi^2+4d}{\pi^2+4t}\,.
\end{equation}
\subsubsection*{Evaluation of $r_k\,\,(k\neq0)$  in the O(4) Case}
\begin{equation}
T_k=\frac{1}{H^2}\int_0^\pi\dfrac{d\eta}{\sin^4\eta}\sin^3\eta\,(A_k^*P+B_k^*Q)(A_kP+B_kQ)
\end{equation}
Here we have suppressed the indices and arguments on the Legendre functions $P$ and $Q$. We substitute $-\cos\eta=x\Rightarrow\sin\eta\,d\eta=dx$ and $\dfrac{d\eta}{\sin\eta}=\dfrac{dx}{1-x^2}$. We then get 
$$T_k=\frac{1}{H^2}(|A_k|^2T_k^{(1)}+(A_k^*B_k+B_k^*A_k)T_k^{(2)}+|B_k|^2T_k^{(3)}),$$ 
where
\begin{eqnarray}
T_k^{(1)}&=&\int_{-1}^1\frac{dx}{1-x^2}(P^{3/2}_{k+1/2}(x))^2\\
T_k^{(2)}&=&\int_{-1}^1\frac{dx}{1-x^2}P^{3/2}_{k+1/2}(x)\,Q^{3/2}_{k+1/2}(x)\\
T_k^{(3)}&=&\int_{-1}^1\frac{dx}{1-x^2}(Q^{3/2}_{k+1/2}(x))^2.
\end{eqnarray}
Similarly $D_k=\dfrac{(-1)^k}{H^2}(A_k^2D_k^{(1)}+2A_kB_kD_k^{(2)}+B_k^2D_k^{(3)})$ with $D_k^{(i)}=T_k^{(i)}$.\\\\
From the definitions of the associated Legendre functions we have\footnote{We will write $F$ instead of $_2F_1$.}:
\begin{eqnarray}
P_{k+1/2}^{3/2}(x)&=&\bigg(\frac{1+x}{1-x}\bigg)^{3/4}\frac{F(-k-1/2,k+3/2,-1/2;(1-x)/2)}{\Gamma(-1/2)}\\
Q_{k+1/2}^{3/2}(x)&=&\frac{\pi}{2}k(k+1)(k+2)\bigg(\frac{1-x}{1+x}\bigg)^{3/4}\frac{F(-k-1/2,k+3/2,5/2;(1-x)/2)}{\Gamma(5/2)}.\nonumber\\
\end{eqnarray}
The above integrals become 
\begin{eqnarray}
T_k^{(1)}&=&\frac{1}{(\Gamma(-1/2))^2}\int_{-1}^1dx\,\frac{(1+x)^{1/2}}{(1-x)^{5/2}}\,F^2(-k-1/2,k+3/2,-1/2;(1-x)/2)\\
T_k^{(2)}&=&\frac{\pi k(k+1)(k+2)}{2\Gamma(-1/2)\Gamma(5/2)}\int_{-1}^1dx\,F(-k-1/2,k+3/2,-1/2;(1-x)/2)\\
&\times&\,F(-k-1/2,k+3/2,5/2;(1-x)/2)\nonumber\\
T_k^{(3)}&=&\frac{(\pi k(k+1)(k+2))^2}{(2\Gamma(5/2))^2}\int_{-1}^1dx\,\frac{(1-x)^{1/2}}{(1+x)^{5/2}}\,F^2(-k-1/2,k+3/2,5/2;(1-x)/2).
\end{eqnarray}
All of the above integrals are divergent. However it turns out that the ratios $T_k^{(2)}/T_k^{(1)},\,T_k^{(3)}/T_k^{(1)}\rightarrow0$, therefore we have 
\begin{eqnarray}
r_k=(-1)^k\frac{A_k^2}{|A_k|^2}=e^{i(\arg A_k+k\pi)} ,
\end{eqnarray}
whence we find that $|r_k|=1$.

\section*{Appendix C: Mode Comparison to non-Fock Modes}
\subsection*{Transformation between the non-Fock de Sitter Invariant Modes of \cite{Kirsten:1993ug} and the SJ Modes}
\label{kgtrans}
The PJ function in terms of the modes we use in this appendix, is
\begin{equation}
    i\Delta (x,x')=i\frac{H^2}{2} \left(f(x)-f(x')\right)+\sum_q u_q(x)u_q^*(x')-u_q^*(x)u_q(x),
\end{equation}
where $f(x)=\eta_x-\frac{1}{2}\sin 2\eta_x-\frac{\pi}{2}$, and for simplicity $q$ refers to the principle index and we will omit the angular indices. The SJ modes then are
\begin{equation}
    s_k(x)=\frac{1}{\lambda_k}\langle i\Delta(x,x') ,s_k(x')\rangle=\sum_q\left(u_q(x) A_{qk}+u_q^*(x)B_{qk}\right)+i\frac{H^2}{2}C_k+i\frac{H^2}{2} f(x) D_k,
    \label{skkg}
\end{equation}
where $u_q$ are the $O(4)$ modes and $A_{qk}=\frac{1}{\lambda_k}\langle u_q,s_k\rangle$, $B_{qk}=-\frac{1}{\lambda_k}\langle u_q^*,s_k\rangle$, $C_k=\frac{1}{\lambda_k}\langle f,s_k\rangle$, and $D_k=-\frac{1}{\lambda_k}\langle 1,s_k\rangle$.
Using \eqref{skkg} we have the inner products
\begin{equation}
    \frac{1}{\lambda_{k'}}\langle s_k,s_{k'}\rangle=\sum_q \left(A^*_{qk}A_{qk'}-B^*_{qk}B_{qk'}\right)+i\frac{H^2}{2}\left(C^*_kD_{k'}-D^*_kC_{k'}\right)=\delta_{kk'}
    \label{inkg2}
\end{equation}
\begin{equation}
    \frac{1}{\lambda_{k'}}\langle s_k^*,s_{k'}\rangle=\sum_q \left(A_{qk'}B_{qk}-A_{qk}B_{qk'}\right)+i\frac{H^2}{2}\left(D_kC_{k'}-C_kD_{k'}\right)=0.
    \label{inkg3}
\end{equation}
Again using \eqref{skkg} and the definition of the coefficients, we have
\begin{equation}
    A_{qk}=\frac{1}{\lambda_k}\sum_{n\neq 0}\left(\langle u_q, u_n\rangle A_{nk}+\langle u_q^*, u_n\rangle^* B_{nk}\right)+i\frac{H^2}{2 \lambda_k}\cancelto{0}{\langle u_q,1\rangle} C_k+i\frac{H^2}{2 \lambda_k}\cancelto{0}{\langle u_q,f\rangle} D_k,
\end{equation}
where the last two inner products vanish because $q\neq 0$ and $\langle Y_q,Y_0\rangle=0$, where the $Y$'s are  spherical harmonics. Similarly,
\begin{equation}
    B_{qk}=-\frac{1}{\lambda_k}\sum_{n\neq 0}\left(\langle u^*_q, u_n\rangle A_{nk}+\langle u^*_q, u_n\rangle^* B_{nk}\right).
\end{equation}
The definitions of $A_{qk}$ and $B_{qk}$ for $q\neq 0$ and $k\neq 0$ are the same as in the $O(4)$ case, and they are therefore ill-defined.
\begin{equation}
C_k=\frac{1}{\lambda_k}\sum_q\left(\cancelto{0}{\langle f, u_q\rangle}A_{qk}+\cancelto{0}{\langle f, u_q^*\rangle}B_{qk}\right)+i\frac{H^2}{2 \lambda_k}\langle f, 1 \rangle+i\frac{H^2}{2 \lambda_k}\langle f, f \rangle
\label{ccoef}
\end{equation}
\begin{equation}
D_k=-i\frac{H^2}{2 \lambda_k}\langle 1, 1 \rangle-i\frac{H^2}{2 \lambda_k}\langle 1, f \rangle.
\label{dcoef}
\end{equation}
Let $C_k=D_k=0$ for $k\neq 0$\footnote{Justification for $C_k=D_k=0$ when $k\neq 0$: If $C_k=\frac{i}{H \alpha_k}e^{i\theta_k}$, $D_k=\frac{\alpha_k}{H}e^{i\theta_k}$, then from \eqref{inkg2} we need that $-i\frac{H^2}{2}\left(D^*_kC_{k'}-C^*_kD_{k'}\right)=e^{i(\theta_{k'}-\theta_k)}\propto\delta_{kk'}$. Therefore we must choose only one special value of $k$ for which $C_k$ and $D_k$ are not 0. From the equation in the second sentence of the next footnote, we see that this special value of $k$ is $k=0$.}, and $A_{qk}=B_{qk}=0$ for $k=0$\footnote{Justification for $A_{qk}=B_{qk}=0$ when $k=0$: Let $A_{qk}, B_{qk}\neq 0$ for some $q$. Then \eqref{inkg2} becomes $A^*_{q0}A_{q0}-B^*_{q0}B_{q0}-i\frac{H^2}{2}\left(C_0D^*_0-C^*_0D_0\right)=1$. But then $\langle s_0,s_q\rangle=\lambda_q\left(A^*_{q0}A_{qq}-B^*_{q0}B_{qq}\right)=0$. This is solved by either a) $A_{q0}=-\sinh\alpha_q e^{-i\beta_q}, B_{q0}=-\cosh\alpha_q$, or b) $A_{q0}=1/\cosh\alpha_q, B_{a0}=e^{i\beta_q}/\sinh \alpha_q$. But neither of these solutions yield vanishing $\langle s_0,s_q^*\rangle$. Therefore we must have $A_{qk}=B_{qk}=0$.}. We can then write
$A_{qk}=\delta_{qk}\cosh{\alpha_k}$, $B_{qk}=\delta_{qk}\sinh{\alpha_k e^{i\beta_k}}$, and \eqref{inkg2}-\eqref{inkg3} become
\begin{equation}
    -i\frac{H^2}{2}\left(D^*_0C_0-C^*_0D_0\right)=1,
    \label{const1}
\end{equation}
\begin{equation}
    C_0D_0-C_0D_0=0.
    \label{const2}
\end{equation}
The constraint \eqref{const2} is trivially satisfied, and \eqref{const1} is satisfied if we choose 
\begin{equation}
C_0=\frac{i}{H\alpha}e^{i\theta}, \indent D_0=\frac{\alpha}{H}e^{i\theta}\indent(\alpha,\theta\in{\rm I\!R}).
\end{equation}
Plugging these into \eqref{ccoef} we get 
\begin{equation}
    \frac{2\lambda_0}{i H^2} \frac{i}{\alpha H}=\langle  f, 1\rangle\frac{i}{\alpha H} e^{i\theta}+\langle f,f\rangle \frac{\alpha}{H} e^{i\theta}.
\end{equation}
$\langle f, 1\rangle$ vanishes, leaving
\begin{equation}
    \alpha^2=\frac{2\lambda_0}{H^2\langle f, f\rangle}.
    \label{alph1}
\end{equation}
Similarly, from \eqref{dcoef} we get 
\begin{equation}
    \alpha^2=\frac{\langle 1, 1\rangle H^2}{2\lambda_0}.
    \label{alph2}
\end{equation}
Together \eqref{alph1} and \eqref{alph2} yield
\begin{equation}
    \alpha^2=\sqrt{\frac{\langle 1, 1\rangle}{\langle f, f\rangle}}=|\text{const}|,
\end{equation}
where $\text{const}$ is a non-zero and finite constant.
Hence $C_0$ and $D_0$ are finite and well-defined.
\section*{Appendix D: Dimensional Analysis}
Dimensional analysis tells us what are the right quantities to compare. 
\subsection*{Dimensional analysis in the continuum}
The retarded Green function satisfies the KG equation \eqref{kgeq} so we have\footnote{$[\,]$ refers to length dimension.} $[G]=2-d=[\Delta]$. The eigenvalue equation for the PJ operator is 
\begin{equation}
(i\Delta\,f_k)(x)=\int dV_y\,i\Delta(x,y)f_k(y)=\lambda_k\,f_k(x).
\end{equation}
Therefore $[\lambda_k]=2$. The SJ two-point function is the positive part of the PJ operator and is given by 
\begin{equation}
W(x,y)=\sum_k\lambda_k\,\tilde f_k(x)\tilde f^*_k(y)\quad\quad(\lambda_k>0),
\end{equation}
where $\tilde f_k$ are the normalised (in the $L^2$ norm) eigenfunctions. So we have, $[\tilde f_k]=-d/2$, $[W]=2-d$.\\\\
\textbf{Note:} If we define the SJ modes as $f_k^{SJ}=\sqrt{\lambda_k}\,\tilde f_k$ then we get $[f_k^{SJ}]=1-d/2$. 
\subsubsection*{Dimensional analysis in the causal set}
The Green function in the causal set is given by \eqref{retgreen},
where $[m^2/\rho]=d-2$ and we can get the dimension of $K_0$ by requiring that $[K_0m^2/\rho]=0$, which yields $[K_0]=2-d=[G]=[i\Delta]$.\\\\
We use the following correspondence to define the analogs of integral operators in the causal set 
\begin{equation}
\int dV_y\rightarrow\frac{1}{\rho}\sum_y\,.
\end{equation}
The eigenvalue equation is given by a matrix equation
\begin{equation}
\frac{1}{\rho}\,i\Delta f_k=\lambda_k\,f_k\,.
\label{csee}
\end{equation}   
Here $[\lambda_k]=2$. These eigenvalues can be compared with the continuum eigenvalues.\footnote{Typically the $\frac{1}{\rho}$ factor in \eqref{csee} is omitted, which is why in the figures above showing the eigenvalues, the causal set spectra are divided by $\rho$.} 
As in the continuum, we have\footnote{This normalisation is obtained by taking the dot product of the vector with itself, divided by the density.} $[\tilde f_k]=-d/2$ and $[f_k^{SJ}]=1-d/2$. For the two-point function we have $[W]=2-d$. Therefore, $W$ can also be compared directly with its counterpart in the continuum. 

\bibliography{references}{}
\bibliographystyle{ieeetr}
\end{document}